\documentclass[12pt,preprint]{emulateapj}

\usepackage{graphicx}
\usepackage[pdftitle={Assesing the procedence of C/2013 U1},
  colorlinks = true, breaklinks = true, citecolor = blue, linkcolor =
  blue, urlcolor = blue, pdfauthor = {Jorge I. Zuluaga}]{hyperref}
\usepackage{natbib}
\usepackage{color}

\newcommand{\sub}[1]{_{\rm #1}}
\renewcommand{\sup}[1]{^{\rm #1}}
\newcommand{\beq}[1]{\begin{equation}\label{#1}}
\newcommand{\eeq}{\end{equation}}
\newcommand{\beqn}{\begin{eqnarray}}
\newcommand{\eeqn}{\end{eqnarray}}

\newcommand{\vrel}{v_{\rm rel}}
\newcommand{\sci}[1]{\times 10^{#1}}

\newcommand\clearrow{\global\let\rowmac\relax}

\newcommand{\hl}[1]{{\color{black} #1}}
\newcommand{\hll}[1]{{\color{black} #1}}

\newcommand{\hb}[1]{{\color{black} #1}}
\newcommand{\hbb}[1]{{\color{red} #1}}

\newcommand{\Oumuamua}{`Oumuamua}
\newcommand{\MUA}{\Oumuamua}

\begin{document}

\title{A general method for assessing the origin of interstellar small bodies:\\the case of 1I/2017 U1 (\Oumuamua)}

\author{Jorge I. Zuluaga, Oscar S\'anchez-Hern\'andez, Mario Sucerquia \& Ignacio Ferr\'in}
\affil{Solar, Earth and Planetary Physics Group \& Computational Physics and Astrophysics Group (FACom)\\ Instituto de F\'\i sica - FCEN, Universidad de Antioquia, Calle 67 No. 53-108, Medell\'in, Colombia}

\begin{abstract}

With the advent of more and deeper sky surveys, the discovery of interstellar small objects entering into the Solar System has been finally possible.  In October 19, 2017, using observations of the \hll{Pan-STARRS} survey, a fast moving object, now officially named 1I/2017 U1 (\Oumuamua), was discovered in a heliocentric unbound trajectory suggesting an interstellar origin. Assessing the provenance of interstellar small objects is key for understanding their distribution, spatial density and the processes responsible for their ejection from planetary systems.  However, their peculiar trajectories place a limit on the number of observations available to determine a precise orbit. As a result, when its position is propagated $\sim 10^5-10^6$ years backward in time, small errors in orbital elements become large uncertainties in position in the interstellar space.  In this paper we present a general method for assigning probabilities to nearby stars of being the parent system of an observed interstellar object.  We describe the method in detail and apply it for assessing the origin of \hb{\Oumuamua}.  A preliminary list of \hl{potential progenitors} and their corresponding probabilities is provided. In the future, when further information about the object and/or the nearby stars be refined, the probabilities computed with our method can be updated.  We provide all the data and codes we developed for this purpose in the form of an open source {\tt C/C++/Python package}, {\bf\tt iWander} which is publicly available at \url{http://github.com/seap-udea/iWander}.
\end{abstract}

\keywords{Interstellar small body  --- asteroids: individual: 1I/2017 U1 --- Methods: numerical}


\section{Introduction}

The detection of interstellar small objects, wandering through the Solar System, has challenged the astronomers over the last century (see e.g. \citealt{Opik1932, McGlynn1989, Sen1993}).  The detection, characterization and investigation of the origin of these objects could shed light on the processes of planetary formation, the architecture of their parent planetary systems, as well as on planetary ejection mechanisms \hl{(see e.g \citealt{Raymond2017})}. Moreover, interstellar visitors can also be the target for future exploration missions (the first ones traveling to an interstellar object; \citealt{Mamajek2017}), become an indirect way to detect planetary systems of nearby stars and to infer some of their properties \citep{Trilling2017} or help to test bold hypotheses about the origin and distribution of life in the Universe, such as lithopanspermia \citep{Adams2005,Belbruno2012}. 

Recently, \citealt{Engelhardt2017} calculated the number density of detectable interstellar objects, obtaining a discouraging value of $1.4 \times 10^{-4}$ au$^{-3}$ which seemed to restrict severely the chances of detecting one of these objects in the coming years, at least with the current available instrumentation.  Nevertheless, on 19 October 2017 a small object with an uncommon orbit ($e=1.1995\pm 0.0002$ and $v_\infty = 26.33\pm 0.03$ km/s) was discovered using the \hll{Pan-STARRS} Survey \citep{Chambers2016}. The orbital properties \hl{determined at that time using} a short orbital arc, while the object was still visible after their rapid passage through perihelion, suggested an interstellar origin.  This result was supported by follow-up spectroscopic observations \citep{Bolin2017} and other theoretical considerations \citep{Ye2017,Masiero2017}. The interloper object has been now named ``1I/2017 U1 (\Oumuamua)'' or \hl{simply} \Oumuamua\ as we will call it consistently through this paper\footnote{IAU Minor Planet Center, 2017, U183}. 

The first follow-up observations suggested that the object was of asteroidal origin, and its composition seemed to be similar to that of transneptunian objects (TNOs) \citep{Merlin2017}.  More recently, \citet{Jewitt2017} measured the observed color and suggested a composition overlapping the mean colors of D-type Trojan asteroids and other inner solar system populations, but inconsistent with the ultra-red matter found in the Kuiper belt. \citet{Ferrin2017} compared the color of the object with other 21 active and extinct Solar System comets and suggest that \MUA\ could actually be an inactive extrasolar comet.  This result could be very important for assessing their dynamical origin.  \hl{On the theoretical side, \citet{Raymond2017} suggested than from a dynamical standpoint it is more likely that \Oumuamua\ be cometary rather than asteroidal in origin.}

Only a few days after the announcement of the discovery, several works attempting to pinpoint its interstellar origin were published in the form of research notes and short papers \citep{Mamajek2017,delafuente2017,Gaidos2017,Portegies2017,Dybczyski2017}.  Most of these initial works were able to restrict the origin of the object to nearby stars and young planetary systems from which it could have escaped via planet-planet scattering.  These early attempts, however, either rely on rough comparisons between the heliocentric entrance velocity of the object and that of nearby stars \citep{Mamajek2017,Gaidos2017} and/or on the numerical propagation of the orbit of the object among many nearby stars, in the galactic potential \hl{\citep{Portegies2017,Dybczyski2017,Feng2018}}. These latter efforts, however, \hl{originally} lack of a rigorous consideration of the orbit uncertainties and the errors in the astrometric and radial velocities of the stars.  \hl{A more rigorous kinematic treatment was recently presented by \citet{Feng2018}.  However, their approach lack of a estimation of the probability that the object were originated in any of their potential progenitors.}  Several authors have even suggested that given the current orbital uncertainties and astrometric information about nearby stars, pinpointing the exact origin of \MUA\ (and possibly other future interstellar objects) to a specific stellar system could be impossible \citep{Zhang2018}.

The detection of \MUA\ however, allowed some authors to update estimations of the number densities of interstellar objects and their detection probability. Thus, for instance, \citealt{Trilling2017} recently determined that if during planet formation processes the ejected mass was about 20 Earth's masses, the detection rate of such interstellar objects could be at least 0.2 per year with the current instrumentation.  This result is consistent with the discovery of \MUA. Moreover, when the Large Synoptic Survey Telescope (LLST, \citealt{Ivezic2008}) begins its wide, fast and deep all-sky survey, the rate could climb to 1 object per year.  As a result, developing a general and \hl{kinematically} rigorous strategy for assessing the origin of this and probably other objects detected in the future, is an important goal to pursue. 

The problem of tracking small Earth's impactors towards their progenitor objects in the Solar System has been extensively studied in the literature (see e.g. \citealt{Strom2005,Zuluaga2013,Strom2015,Zuluaga2018}).
However, the dynamic of the Solar System is well constrained, the timescales are relatively short and the the position of the perturbing objects \hl{are} known to exquisite levels of precision.  This is not the case when reconstructing the orbit of an interstellar object. Timescales are huge, amplifying even small uncertainties in initial positions.  The perturbing objects (stars) are very distant and small errors in their projected position in the sky correspond to large errors in their position in space.  \hl{Moreover, the gravitational scattering from random encounters with nearby stars, may prevent a successful track-down of their origins.}  \hb{However,} with the advent of \hl{new} precise astrometric \hl{data} such as \hl{that provided by} Gaia \hb{and the fact that we are now aware that a non-negligible number of detectable interstellar interlopers, could be wandering through the solar system}, the situation is improving significantly. 

Tracking the motion of objects in the interstellar space is not new.  Several authors have studied the problem, aiming to determine the past and future close encounters of the Sun with nearby stars (See e.g. \citealt{Garcia2001,Bailer2015,Berski2016}). Of particular interest for this work is the recent paper by \citet{Bailer2017} which use up-to-date astrometric and radial velocity catalogs to study this problem, while rigorously considering the uncertainties involved. \hl{The present} paper is inspired and mostly based on the models, data and techniques developed in that work.

In this paper we present a general methodology and a computer tool designed for assessing the \hl{kinematic} origin of an interstellar object.  Instead of determining which particular stellar system could be the source of a given object, the methodology presented here aims at computing ``interstellar origin probabilities'' (hereafter IOP) of many nearby stars. For that purpose \hl{our work} takes into account rigorously the uncertainties in the object and stars kinematic properties. When more and better astrometric information become available, the \hl{values of the} IOP \hl{computed with our methods and software} can be improved.  

This paper is organized as follows: \hb{we introduce the basics of our probability model and define the concept of ``Interstellar Origin Probability'' (IOP) in \autoref{sec:ProbabilityModel}.}  In \autoref{sec:Outline} we outline the method \hb{and tools used in this work to estimate probabilities}.  Sections \ref{sec:SurrogateObjects}-\ref{sec:InterstellarMotion} present the details of the implementation, while illustrating each technique in the particular case of \MUA.  Section \ref{sec:IOP} defines \hb{rigorously each of the terms in the calculation of the IOP}.  The results of applying the methodology to the case of \MUA\ are presented in \autoref{sec:Results} including a list of the potential progenitors with their respective IOP.  Section \ref{sec:Discussion} is devoted at discussing the limitations of the method, but also its future potential as a general framework to study the orbit of \hl{newly} discovered \hl{interstellar} objects.  Finally, in \autoref{sec:SummaryConclusions} we present a summary of the paper and draw the main conclusions derived from it.

\section{Probability model}
\label{sec:ProbabilityModel}

\begin{figure*}[t]
{\centering
\includegraphics [width=140mm] {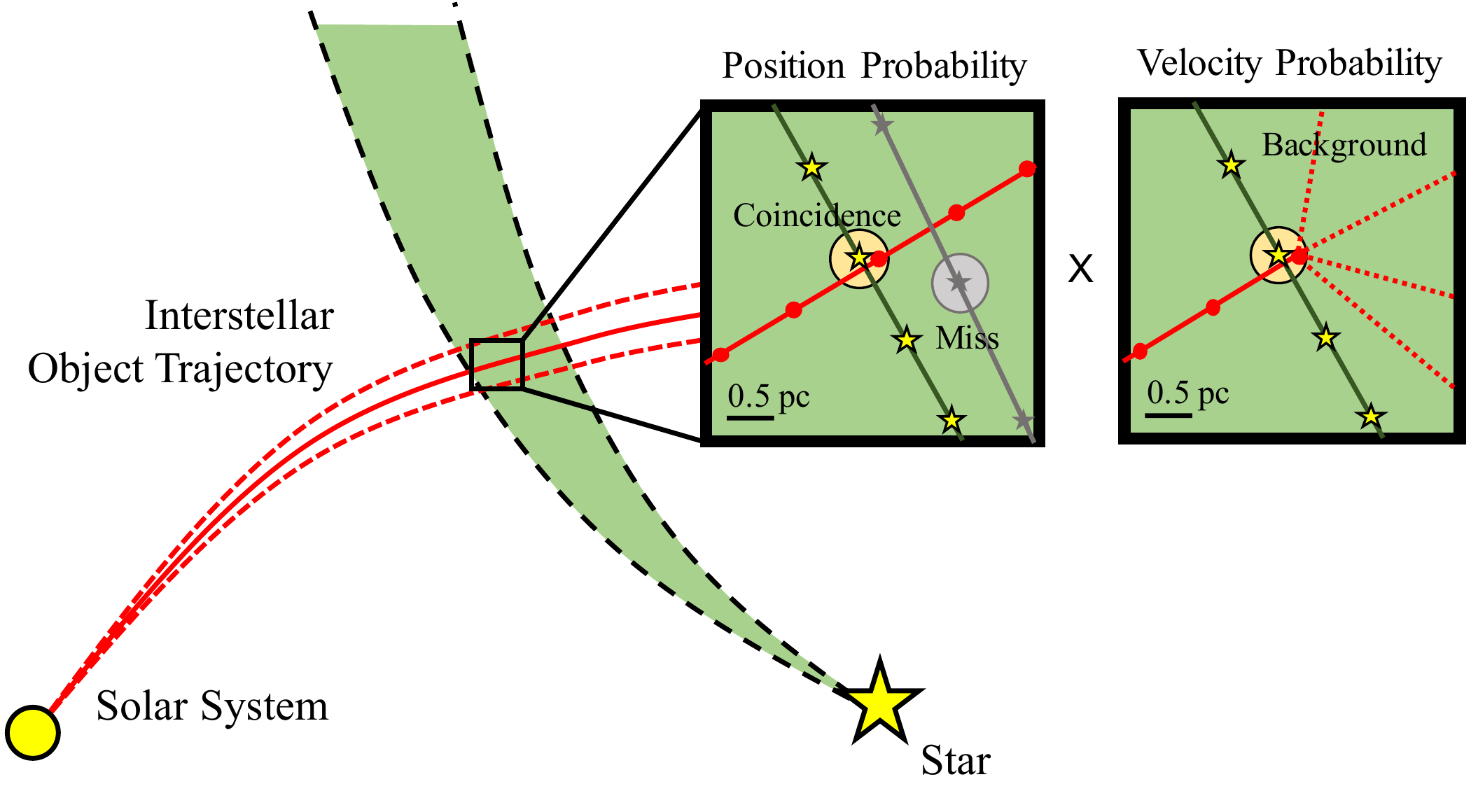}
\caption{\hbb{Schematic representation of the probability model
    devised in this model to calculate the Interstellar Origin
    Probability (IOP).  Given the uncertainty in the orbit of the
    interstellar object (ISO) and a given star, we can imagine a large
    number of parallel universes where the trajectories of both, the
    ISO and the star, either coincide in space and time (coincidence)
    or not (miss).  The ratio of the number of those parallel
    universes where the trajectories coincide give us the {\it
      position probability}. Even in the case of coincidence, the
    object may come from the background (dashed lines in the rightmost
    box) instead of being an indigenous object. The IOP is the product
    of the position probability and a correcting factor (the veloicity
    probability) taking into account the flux of background
    objects}.\medskip
\label{fig:ProbabilityModel}
}}
\end{figure*}

\hb{

In this work we aim at assessing the following question: 

\begin{quotation}
\noindent \em If you observe an interstellar object coming towards the Solar System from a given direction in space, what is the probability (origin probability) that a specific star be the progenitor of the object?
\end{quotation}

With the exception of trivial cases, defining a mathematical model for computing probabilities is tricky.  You need to properly define your probability space \citep{Kolmogorov1968}, including, constructing a probability function that connects ``events'' with real numbers quantifying the ``frequency'' of those events.  What is the probability space in this case? how can we define probabilities in the context of our problem?.  In \autoref{fig:ProbabilityModel} we schematically depict the way as the ``origin probability'' is defined in the context of this work.

The uncertainty in the orbit of the interstellar object and the astrometric parameters of the star (position and proper velocity), makes that we cannot define a single trajectory for both. Instead, we have a beam of trajectories that eventually intersect in a relatively small volume in the past (square region in  \autoref{fig:ProbabilityModel}).  

To estimate probabilities, we can imagine that a large number of ``parallel'' universes exists where the object and the star have one of many particular set of orbital properties inside their respective trajectory beams.  In some of those universes, the trajectories may coincide within a given critical radius (eg. the truncation radius of the planetary system, see next section).  In others, the properties of the object and the star are such that their trajectories, although reach a minimum distance at some point in the past, have a encounter distance large enough to prevent us thinking that the object could be ejected from the star (we call this condition a ``miss'').

Under a frequentist approach, the probability of coincidence in position (hereafter ``position probability'') will be the ratio of the number of universes in which we have a coincidence, $N\sub{pos}$, to the number of all possible trajectory configurations.

However, among all those universes where there is a coincidence in position, we cannot ensure that the interstellar object was actually ejected from the stellar system (an ``indigenous'' object).  It could happen that the object be simply the result of the gravitational scattering of a background object by the stellar system (dashed trajectories in the leftmost panel of \autoref{fig:ProbabilityModel}). 

How can we distinguish these two cases?.  Since ejection and scattering are different physical phenomena, it is reasonable to expect that the distribution of speeds for indigenous objects be different than that of background objects.  Thus for instance, and as we will show in \autoref{subsec:RelativeVelocity}, while the mechanism ejecting indigenous objects tend to produce Maxwellian-like speed distributions, which are characterized by the fact that low and high speeds have low probabilities, the speeds of scattered background objects arises from a combination of ejection and the relative speeds of stars in the Galaxy; for simplicity we can be assumed that background relative speed are nearly uniform, ie. low and large speeds have similar probabilities.

If we assume that $p\sub{ind}(\vrel) d\vrel d\Omega$ is the number of the total number indigenous objects ejected from the stellar systems with relative speeds between $\vrel$ and $\vrel+d\vrel$ in the direction of the Solar System, and $p_B(\vrel)d\vrel d\Omega$ is the number of background objects in the same speed interval, the fraction of the number of spatial coincidences that actually corresponds to an ejection process, can be estimated as:

\beq{eq:find_base}
f\sub{ind}(\vrel) = \frac{p\sub{ind}(\vrel)}{p_i(\vrel)+p_B(\vrel)}
\eeq

Here, two different extreme situations can be considered:

\begin{itemize}

\item[(i)] The number of indigenous objects are much larger than that of background objects, ie. $p\sub{ind}(\vrel)\gg p_B(\vrel)$. If this is the case, 

$$
f\sub{ind}(\vrel)\approx 1
$$

This will be the case, for instance, if we have a young stellar system with an ongoing planetary formation process.

\item[(ii)] The number of background objects coming out from the stellar system are much larger than the number of indigenous ones, ie. $p_B(\vrel)\gg p\sub{ind}(\vrel)$.  In this case,

$$
f\sub{ind}(\vrel)\approx \frac{p\sub{ind}(\vrel)}{p_B(\vrel)}
$$

This will be the case of moderately old stellar system where scattering still exists but at a more moderate rate.

\end{itemize}

Although demonstrating which is the case for a particular star is out of the scope of this work, we will assume that the second situation is more common.  Despite this particular assumption, in our detailed model we will still include enough information to estimate origin probabilities for both extreme cases.

In summary, the probability that a given star be the progenitor of an interstellar object, namely its ``interstellar origin probability'' (IOP), is proportional to:

\beq{eq:IOP}
{\rm IOP}\propto N\sub{pos} f\sub{ind}
\eeq

In the following paragraphs we first summarize the numerical procedure we used to set up our ``probability space'' and then the detail mathematical models used to estimate the factors $N\sub{pos}$ and $f\sub{ind}$, required to calculate the IOP of a sample of nearby stars.
}

\section{Outline of the Method}
\label{sec:Outline}

Once a small-body is discovered in the Solar System, \hl{the astrometry and a reference orbit (computed from the observation weighting differences) is published by the Minor Planet Center (MPC). An up-to-date best-fit orbital solution of the astrometric data provided by the MPC is then computed and made available publicly in the} {\tt JPL Small-Body Database Browser}\footnote{\url{https://ssd.jpl.nasa.gov/}}.  The JPL orbit solution is provided in the form of a vector of ``nominal orbital elements'' $E_0:(e_0,q_0,t_{p,0},\Omega_0,\omega_0,i_0)$ (orbital eccentricity, perihelion distance, time of perihelion passage, longitude of the ascending node, argument of the perihelion and inclination, respectively) at a reference epoch $t_0$. Along with this information, the \hl{JPL} also provides the nominal mean orbital motion $n_0$, and its standard error $\sigma_{n0}$. The uncertainties in the orbital fit are characterized with an ``orbit covariance matrix'', $C_{jk}$, defined as:

\beq{eq:Covariance}
C_{jk}=\left\{
\begin{array}{cc}
\sigma_j^2 & j=k\\
\rho_{jk}\sigma_j\sigma_k & j\neq k
\end{array}
\right.
\eeq

where $j:e,q,t_p,\Omega,\omega,i$ and $\rho_{jk}$ are the correlations among the orbital elements.

\hb{Provided the nominal orbit and its associated errors, we need to propagate backward in time the trajectory of the object and the position of a set of nearby stars to set up our probability space and prepare the conditions to compute the IOP. The specific procedure we follow to achieve this goal is summarized as follow:}

\begin{enumerate}

\item Generate $N_p$ clones of the orbital elements vector $E$ \hl{of the interstellar object} compatible with the latest orbital solution $E_o$ and $C_{jk}$ (see \autoref{sec:SurrogateObjects}).  These clones describe the orbit of what we will call the ``surrogate objects'' (following the convention of \citealt{Bailer2017}).

\item Integrate backwards the orbits of the surrogate objects, until a time when the object in the nominal orbit reaches a distance from the Sun where the galactic tides become relevant for the dynamic of the object (see \autoref{sec:SolarSystemIntegration}).

\item Using the ``linear motion approximation'' (LMA), identify the stars in an astrometric and radial velocity catalog, such that their minimum distance to the nominal object, $d\sub{LMA,min}$, be less than or equal to a threshold distance.  Stars identified with this procedure are called the ``candidates'' (see \autoref{sec:AstroRV}).

\item For each candidate, compute a more precise encounter time $t\sub{min}$ and minimum distance $d\sub{min}$ by integrating the orbit of the star and the nominal object in the galactic potential (see \autoref{sec:InterstellarMotion}). 

\item For each candidate star, generate $N_s$ ``surrogate stars'' (hypothetical stars having astrometric properties compatible with the observed properties of the star and their errors). \hb{The orbital integration of both, the surrogate stars and the surrogate object, define the intersecting beam of trajectories in \autoref{fig:ProbabilityModel}}.

\item Identify those candidates for which the more precisely computed minimum distances is below a tighter distance threshold and the IOP probability is the highest.  We call these the ``potential progenitors''.

\end{enumerate}

All the potential progenitors identified with this procedure are tabulated in descending order of IOP probability (Table 4).  The idea is not to select an individual star, but to assign a probability to each of them that can be improved with this method as the \hl{interstellar object} and the stars itself are better known.

\section{The surrogate objects} 
\label{sec:SurrogateObjects}

Since the trajectory of the object is uncertain, and small errors inside the Solar System amplify when the orbit is integrated into the interstellar space, \hl{assessing its interstellar origin require than more than a single orbit (the nominal solution) be computed}. 

\hl{For this purpose} we generate $N_p$ vectors of orbital elements $E$ compatible with the orbital solution described with $E_0$ and $C_{jk}$ (see \autoref{sec:Outline}).  \hl{Random realizations of the orbital elements are computed} using a multivariate Gaussian random number generator\footnote{Most numerical libraries are provided with a multivariate random number generator.  In the particular case of this work we use the generator and related routines provided by the GNU Scientific Library {\tt GSL} \citep{Galassi2002}} having mean values equal to the nominal solution element vector $E_o$ and covariance equal to the orbit covariance published by JPL.

\hl{In \autoref{fig:Radiant} we show the ``ingress'' position and velocity (see next section) of 1000 surrogate objects whose orbits were generated using this procedure. The strong correlation between the orbital parameters places the orbits of the surrogate objects in a very narrow ellipsoid which is projected in the sky at the time of ingress to the Solar System as a narrow line around the nominal radiant of the object (upper panel in \autoref{fig:Radiant}). The error in the present \Oumuamua's orbital solution (JPL 15), propagates as an error in the position at the time of ingress (
$\sim$\hl{2000}
years before present) of 
$\sim$\hl{7} AU 
(middle panel in \autoref{fig:Radiant}), and \hb{a dispersion} in \hb{their velocities} of 
$\sim$ \hl{0.03} km/s
(lower panel in \autoref{fig:Radiant}).}

\medskip
\section{Trajectory in the Solar System}
\label{sec:SolarSystemIntegration}

Once the orbital elements of the surrogate objects have been generated at the reference epoch, we proceed to calculate the trajectory of each object in the gravitational field created by the sun and the 8 planets\footnote{Although dwarf planets and major asteroids were not included in the integrations presented here, it is easy to add them in the {\tt iWander} package provided with this work.}.

For that purpose we use a Gragg-Bulirsch-Stoer algorithm \citep{Gragg1965,Bulirsch1966} adapted from site\footnote{\url{http://www.mymathlib.com/}}. Positions and velocities of the the planets were not computed with the integrator itself, but \hl{using {\tt NASA NAIF SPICE} software (\citealt{Acton1996} and Jon D. Giorgini)\footnote{\url{http://naif.jpl.nasa.gov/pub/naif}}}.  \hl{For that purpose we use} the latest {\tt DE431} planetary kernel. We verified that for the case of \MUA\ our integrations were close to the precise solutions computed by the JPL Solar System Dynamics group and published in the on-line {\tt NASA Horizons} system. \hl{A maximum error} of 0.1\% at back to 400 years before the reference epoch \hl{were obtained with our integrator}.

Since the trajectory of the object is hyperbolic, and in the case of \MUA, highly inclined with respect to the ecliptic (i=122.6$^\circ$), the object reached large distances to the Sun and the planets in relatively short times.  Given the original uncertainty in the orbit, using the integrator to compute a very precise position of the surrogate objects within the Solar System is pointless.  At some time in the past, the accumulated errors in the state vector $(x,y,z,v_x,v_y,v_z)$ due only to the orbit uncertainty, will be much larger than the errors from assuming that the objects move in an ideal hyperbola.

Thus, for instance, in the case of \MUA, we verified that at $t=-100$ years, the positions of the surrogate objects were spread inside an ellipsoid with a characteristic size of 
\hl{$\sim$ 0.5} AU.  
On the other hand, the osculating elements of the object orbit at 
\hl{t=-6 years}, 
when it was at 
$\sim$\hl{40 AU} 
from the Sun and above the ecliptic plane, allows us to predict the position at $t=-100$ years with an uncertainty of 
$\sim$\hl{0.04 AU}.  
We call the orbital elements at this point, the ``asymptotic elements'' and use them to compute the position of the object \hl{inside the Solar System} in the far past.  The value of the asymptotic elements for \MUA\   \hl{along with other relevant information about its orbit in the Solar System} are presented in \autoref{tab:AsymptoticElements}.

Using the asymptotic orbital elements we calculate the position and velocity of the surrogate objects at arbitrary times in the past.  However, when the object is at a distance comparable with the truncation tidal radius of the Solar System, $d\sub{T,\odot}$, the effects of the galactic potential in its motion cannot be neglected.  We call this epoch the ``time of ingress'' $t\sub{ing}$.  The interstellar orbit integration for the surrogate objects and stars\hl{, starts precisely at} $t\sub{ing}$.  

\hl{It is important to stress that the particular value assumed for $d\sub{T,\odot}$ does not modify considerably our final results and the conclusions of this work. In a numerical experiment we find that changing the value of $d\sub{T,\odot}$ from 10,000 AU to 100,000 AU introduces differences below 0.1\% in the interstellar positions and velocities of the surrogate objects, an error which is much smaller than their intrinsic orbital uncertainties.}

In \autoref{tab:AsymptoticElements} we present the time of ingress for \MUA\ and its position and velocity in the ICRS galactic reference frame at that time.  Our results are in agreement to those of \citet{Mamajek2017}.

\begin{table}
  \centering
  \scriptsize
  \begin{tabular}{ll}
  \hline Property & Value \\\hline\hline
  Reference Epoch & 2458059.5 TDB = 2017-Nov-02.0\\
  Nominal elements (JPL 15)
                      & $q = 0.255343194$ AU \\
                      & $e = 1.199512420$ \\
                      & $i = 122.687205$ deg \\
                      & $\Omega = 24.599211$ deg \\
                      & $\omega = 241.702983$ deg \\
                      & $M = 36.425313$ deg \\
                      & $\mu = 1.327124400\sci{11}$ km$^3$/s$^2$\\
  Epoch of Asymptotic elements & 2455736.51 TDB = JUN 24.01 2011\\
  Asymptotic elements & $q = 0.252440118$ AU \\
                      & $e = 1.197253807$ \\
                      & $i = 122.735691$ deg \\
                      & $\Omega = 24.252921$ deg \\
                      & $\omega = 241.680101$ deg \\
                      & $M = -1544.878492$ deg \\
                      & $\mu = 1.327124400\sci{11}$ km$^3$/s$^2$\\
  Asymptotic covariance
                      & Eccentricity \\
  $\times 10^{-6}$   & $C_{ee}$ = 0.028\\
                      & $C_{eq}$ = 0.010\\
                      & $C_{et}$ = 7.558\\
                      & $C_{e\Omega}$ = -0.047\\
                      & $C_{e\omega}$ = 1.905\\
                      & $C_{ei}$ = 1.002\\

                      & Perihelion distance \\
                      & $C_{qq}$ = 0.004\\
                      & $C_{et}$ = 2.831\\
                      & $C_{q\Omega}$ = -0.018\\
                      & $C_{q\omega}$ = 0.714\\
                      & $C_{qi}$ = 0.375\\

                      & Periapsis time \\
                      & $C_{tt}$ = 4847.133\\
                      & $C_{t\Omega}$ = -14.358\\
                      & $C_{t\omega}$ = 514.030\\
                      & $C_{ti}$ = 271.891\\

                      & Long. ascending node \\
                      & $C_{\Omega\Omega}$ = 0.080\\
                      & $C_{\Omega\omega}$ = -3.205\\
                      & $C_{\Omega i}$ = -1.686\\

                      & Argument of periapsis\\
                      & $C_{\omega\omega}$ = 130.021\\
                      & $C_{\omega i}$ = 68.363\\

                      & Inclination\\
                      & $C_{ii}$ = 35.963\\

  Truncation radius & $50000.0$ AU \\
  Time of ingress & $8994.0$ years \\
  Radiant at ingress & RA = $279.80\pm 0.03$ deg\\
                     & DEC = $33.99\pm 0.01$ deg \\
                     & l =  $62.90\pm 0.02$ deg \\
                     & b = $17.11\pm 0.02$ deg \\
  Velocity at ingress & U = $-11.463\pm 0.011$ km/s\\
                      & V = $-22.401\pm 0.001$ km/s\\
                      & W = $-7.748\pm 0.010$ km/s\\\hline
  \end{tabular}
\caption{Properties of the \Oumuamua's orbit in the Solar System.\label{tab:AsymptoticElements}}
\end{table}

\begin{figure}
{
\centering
\includegraphics [width=90mm] {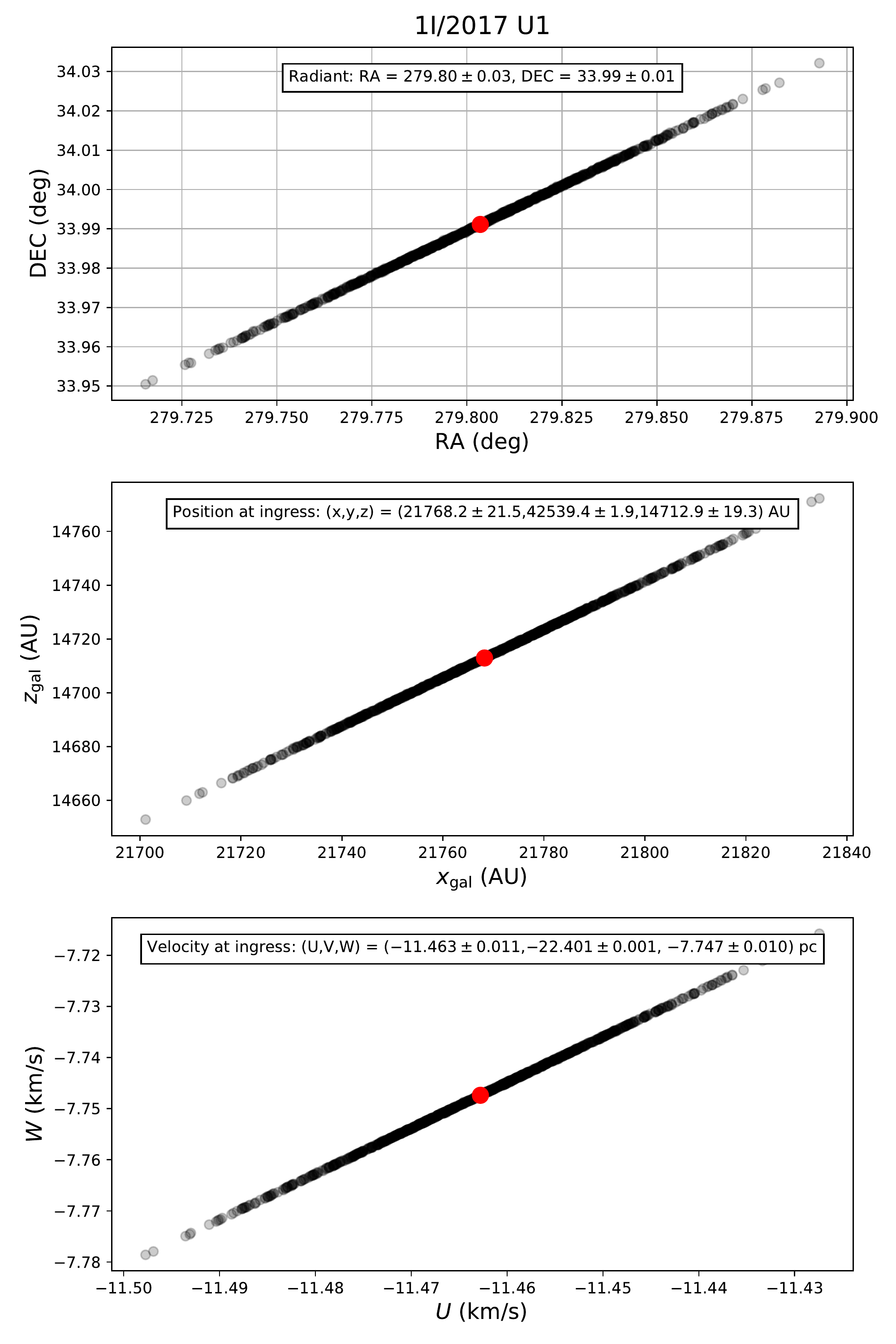}
\caption{Dispersion in the position of the surrogate objects at the time of ingress into the Solar System.  Right ascension, RA and declinations, DEC are referred to the Solar System Barycenter and the J2000 equinox.  Galactic cartesian coordinates ($x\sub{gal},y\sub{gal},z\sub{gal}$) and velocities ($U,V,W$) are referred to the ICRS galactic system of reference.\medskip
\label{fig:Radiant}
}
}
\end{figure}

In order to illustrate the effect that uncertainties in the orbit solution has in the predicted position of the object in the past, we show in \autoref{fig:Radiant} the radiant in the sky of the surrogate objects at ingress time.  We also plot the position in space of the objects in the ICRS galactic reference frame, their velocities in the same reference frame and their corresponding errors.  The convention stands that $(U,V,W)$ correspond to ($v\sub{gal,x},v\sub{gal,y},v\sub{gal,z}$).

As commented before, even a small error in the orbital elements at the reference epoch are propagated as relatively large errors at the ingress time.  While at $t=-100$ yrs the surrogate objects were contained in an ellipsoid with a characteristic size of
\hl{0.5} AU, 
at $t\sub{ing}$ the surrogate objects have spread as a 
$\sim$\hl{100} AU 
cloud \hl{(at this time the objects are by definition at a distance equivalent to the truncation radius, namely
\hl{50,000 AU}
)}.  We have verified that this trend continues into the interstellar space, with a expansion velocity of 
$\sim$\hl{0.05} km/s 
(the ``size'' of the cloud in the velocity $UVW$ space, see bottom panel in \autoref{fig:Radiant}).  At this rate, the cloud characteristic size will grow to 
$\sim$2 pc (the typical distance between stars in the solar neighborhood) in 
$\sim$\hl{40} Myr.  
Beyond this time, \hl{the cloud of surrogate objects will encounter at the same time more than one nearby star}, and hence the reliability of our method will be compromised.  We call this the ``maximum retrospective time'', $t\sub{ret}$.  In the case of \MUA, $t\sub{ret}\sim 40$ Myr.

\section{Astrometric and radial velocity databases}
\label{sec:AstroRV}

In order to compare the position of our interstellar object with the position of nearby stars in the far past, we need to know as precisely as possible the position and velocities of those stars at present time. For this purpose we have compiled up-to-date astrometric information (position, parallaxes, proper motion and radial velocities) of \hl{285,114} stars (see \autoref{tab:AstroRV}).

Compiling precise astrometric and radial velocity measurements for a significant number of nearby stars, is tricky.  On one hand, precise astrometric databases such as Hipparcos \citep{Perryman1997} and Tycho-2 \citep{Hog2000} does not include radial velocities measurements.  The Gaia mission have the potential to provide this information, but it will only be available since Data Release 2 (DR2) in April 2018\footnote{\url{https://www.cosmos.esa.int/web/gaia/release}}.  

On the other hand, there are several public catalogs that provide precise radial velocities (see \autoref{tab:AstroRV}).  However, not all the objects in those catalogs are included in the astrometric catalogs, and in some cases they have unique identifications without any reference to the Hipparcos/Tycho-2 ids, which are to the date of writing this paper, the identification for objects having precise astrometric measurements (either from the Hipparcos or the Gaia mission).

We compile the information required for this work following the detailed directions recently published by \citealt{Bailer2017}.  Additionally, and in order to include nearby bright stars (which were not included in the Gaia catalog) we search for other information about the Hipparcos stars in the {\tt Simbad} information system\footnote{\url{http://simbad.u-strasbg.fr/simbad/}}.

The importance of having this information in a \hl{properly-structured table}, led us to create the so-called {\tt AstroRV} catalog (astrometric and radial velocities catalog).  The catalog is provided with the {\tt iWander} package developed for this work.

For the sake of completeness and reproducibility we outline below the  procedure required to compile the {\tt AstroRV} catalog:

\begin{enumerate}

\item Get available astrometric and radial velocities catalogs.  In \autoref{tab:AstroRV} we summarize the information of the catalogs we use for this work.  The most important one is the Gaia DR1 catalog containing \hl{over 1 billion objects \citep{Brown2016}.  Among those objects, however, only 2,026,210 were previously included in Hipparcos and Tycho-2 catalogs and a full astrometric solution (with a positive parallax) was obtained (Gaia TGAS catalogue). Among them, 93,398 are in the Hipparcos catalog and 1,932,812 belongs to the Tycho-2 catalog}.

\item Obtain the general information available in {\it Simbad} for all the stars having an Hipparcos ID.  The information includes other designations for the stars (Henry Draper catalog id and proper names for the bright stars) and radial velocities for some of those stars. 

\item Find all the objects in the  Hipparcos and Tycho Catalogues \citep{ESA1997} and in the Simbad table compiled before, that are not included in the GAIA \hl{TGAS} catalog.  Append the resulting objects to the latest catalog to create a final table with all the available astrometric information.  

\item Create a table including all objects in the radial velocities catalogs that have Hipparcos and/or Tycho-2 ids.  For those objects not having any of those ids, find the Hipparcos/Tycho-2 objects matching the coordinates within a 50 arcsec radius (we use for this purpose the X-match tool of {\tt Simbad}). For detailed instructions about how to compile \hl{the available} radial velocity \hl{catalogs} please refer to \citet{Bailer2017}.

\item Merge the astrometric and radial velocity tables according to the Hipparcos/Tycho-2 ids.

\end{enumerate}

The catalog compiled with this procedure contains \hl{285,114} stars including the nearest and brightest ones.  We provide with the {\tt iWander} package the required scripts for creating the catalog as described before.  Those scripts can be modified to include future radial velocity and astrometric catalogs, or to update the information in older versions.  An up-to-date version of the {\tt AstroRV} catalog will be available at the {\tt iWander} {\tt GitHub} repository\footnote{\url{http://github.com/seap-udea/iwander.git}}.

\begin{table*}
\centering
\scriptsize
\begin{tabular}{lllllll}
\hline 
Catalog name & Number of objects & Hipparcos ID & Tycho-2 ID & Contribution & CDS Code & Reference \\\hline\hline
\multicolumn{7}{c}{\tt Astrometric} \\
\hline

Gaia TGAS & 2026210 & 93398 & 1932812 & 2026210 & \tt I/337/tgas & (1) \\
Hipparcos & 117955 & 117955 & -- & 24557 & \tt I/239/hip\_main & (2) \\
Tycho & 579054 & -- & 579054 & 164550 & \tt I/259/tyc2 & (2) \\
Simbad & 118004 & 118004 & -- & 67 & \tt -- & (3) \\

  \hline
  {\tt Totals} & \bf 2841223 & \bf 329357 & \bf 2511866 & \bf 2215384 & This work \\
  \hline

  \multicolumn{7}{c}{\tt Radial velocities} \\
  \hline
WEB1995 & 1167 & 494 & 673 & 252 & \tt III/213 & (5) \\
GCS & 14139 & 12977 & -- & 7091 & \tt J/A+A/530/A138 & (6) \\
RAVE-DR5 & 520701 & 121 & 309596 & 217257 & \tt III/279/rave\_dr5 & (7) \\
PULKOVO & 35493 & 35493 & -- & 23412 & \tt III/252/table8 & (8) \\
FAMAEY2005 & 6028 & 6028 & -- & 5544 & \tt J/A+A/430/165/tablea1 & (9) \\
BB2000 & 673 & -- & 673 & 503 & \tt III/213 & (10) \\
MALARODA & 2032 & -- & 2032 & 416 & \tt III/249/catalog & (11) \\
GALAH & 10680 & -- & 10680 & 7837 & \tt J/MNRAS/465/3203 & (12) \\
MALDONADO & 473 & 473 & -- & 301 & \tt J/A+A/521/A12/table1 & (13) \\
APOGEE2 & 29173 & -- & 29173 & 22501 & \tt -- & (14) \\

  \hline
  {\tt AstroRV} & \bf 620559 & \bf 55586 & \bf 352827 & \bf 285114 & This work \\
  \hline

\end{tabular}
\caption{Catalogs used to compile the {\tt AstroRV} catalog for this work. References: (1) \citealt{Brown2016}, (2) \citealt{ESA1997} (3) \citealt{Wenger2000}, (5) \citealt{Barbier2000a}, (6) \citealt{Casagrande2011}, (7) \citealt{Kunder2017}, (8)  \citealt{Gontcharov2006}, (9) \citealt{Famaey2005}, (10) \citealt{Barbier2000b}, (11) \citealt{Malaroda2000}, (12) \citealt{Martell2016}, (13) \citealt{Maldonado2010}, (14) \citealt{Bailer2017}.
\label{tab:AstroRV}}
\end{table*}

\section{Trajectories in interstellar space}
\label{sec:InterstellarMotion}

Having the position and velocity of both, the surrogate objects and the nearby stars, we now may integrate their orbits backward in time to identify any close approach.  Since we need to perform a comparison between the position of at least $\sim$1,000 surrogate objects with ${\cal O}(10^5)$ stars, \hb{simulation} times may become prohibitively large.  In a first step we perform a selection of ``candidate stars'' using the so-called ``linear motion approximation'' (LMA).

\subsection{The LMA approximation}
\label{subsec:LMA}

Under this approximation, all particles, the surrogate objects and the stars, move in straight lines:

$$
\vec{r}_i(t)=\vec{r}_{i,0} + \vec{v}_{i,0} t
$$

\hl{Here, $\vec{r}_i$ ($\vec{r}_{i,0}$) and $\vec{v}_{i}$ are the position and velocity of the $i$-th particle.} The instantaneous distance between \hl{particles $j$ and $k$ (surrogate objects and stars)}, is given by:

\begin{eqnarray}
\label{eq:LMADistance}
\nonumber
|\Delta\vec{r}_{jk}(t)|^2 & = & |\Delta\vec{r}_{jk,0} + \Delta\vec{v}_{jk,0}t|^2 \\
\nonumber
 & = & |\Delta\vec{r}_{jk,0}|^2 + 2 \Delta\vec{r}_{jk,0}\cdot \Delta\vec{v}_{jk,0} t + |\Delta\vec{v}_{jk,0}|^2 t^2
 \end{eqnarray}

where $\Delta\vec{r}_{jk,0}=\vec{r}_{k,0}-\vec{r}_{j,0}$ and $\Delta\vec{v}_{jk,0}=\vec{v}_{k,0}-\vec{v}_{j,0}$.  This function \hl{has a} minimum when $d|\Delta\vec{r}_{jk}(t)|^2/dt=0$, ie. at the time $t\sub{jk,min}$ when the distance between the objects is $d\sub{jk,min}$:

\begin{eqnarray}
\label{eq:LMAtdmin}
t\sub{jk,min} & = & -\frac{\Delta\vec{r}_{jk,0}\cdot\Delta\vec{v}_{jk,0}}{|\Delta\vec{v}_{jk,0}|^2}\\
\nonumber
d\sub{jk, min} & = & |\Delta\vec{r}_{jk}(t\sub{min})|
\end{eqnarray}

We compute $t\sub{jk,min}$ and $d\sub{jk, min}$ for the nominal object and all the stars in our {\tt AstroRV} \hb{catalog} in order to select the \hl{``candidate stars'', namely those stars that could actually have close encounters with the object.}

After numerically testing several criteria, we found that the most cost-effective (in terms of computational resources in the succeeding steps), albeit simple condition is:

\beq{eq:CandidatesCriterion}
d\sub{jk, min}<{\rm sup}\left\{d\sub{max},\frac{1}{5} d_0\right\}
\eeq

\noindent \hl{where $d\sub{max}$ is an arbitrary distance threshold, $d_0$ is the present distance of the star and the $1/5$ is an adjustable numerical factor. For our present analysis we assume $d\sub{max}=10$ pc.}

\hl{It should be stressed that with an arbitrary amount of computational resources all the stars in the {\tt AstroRV} catalog could be integrated in the galactic potential (see next section) and no heuristic selection criterion (such as that in \autoref{eq:CandidatesCriterion}) need to be used.  This criterion is only intended to save computational resources.}

\begin{figure}
{
\centering
\includegraphics [width=90mm] {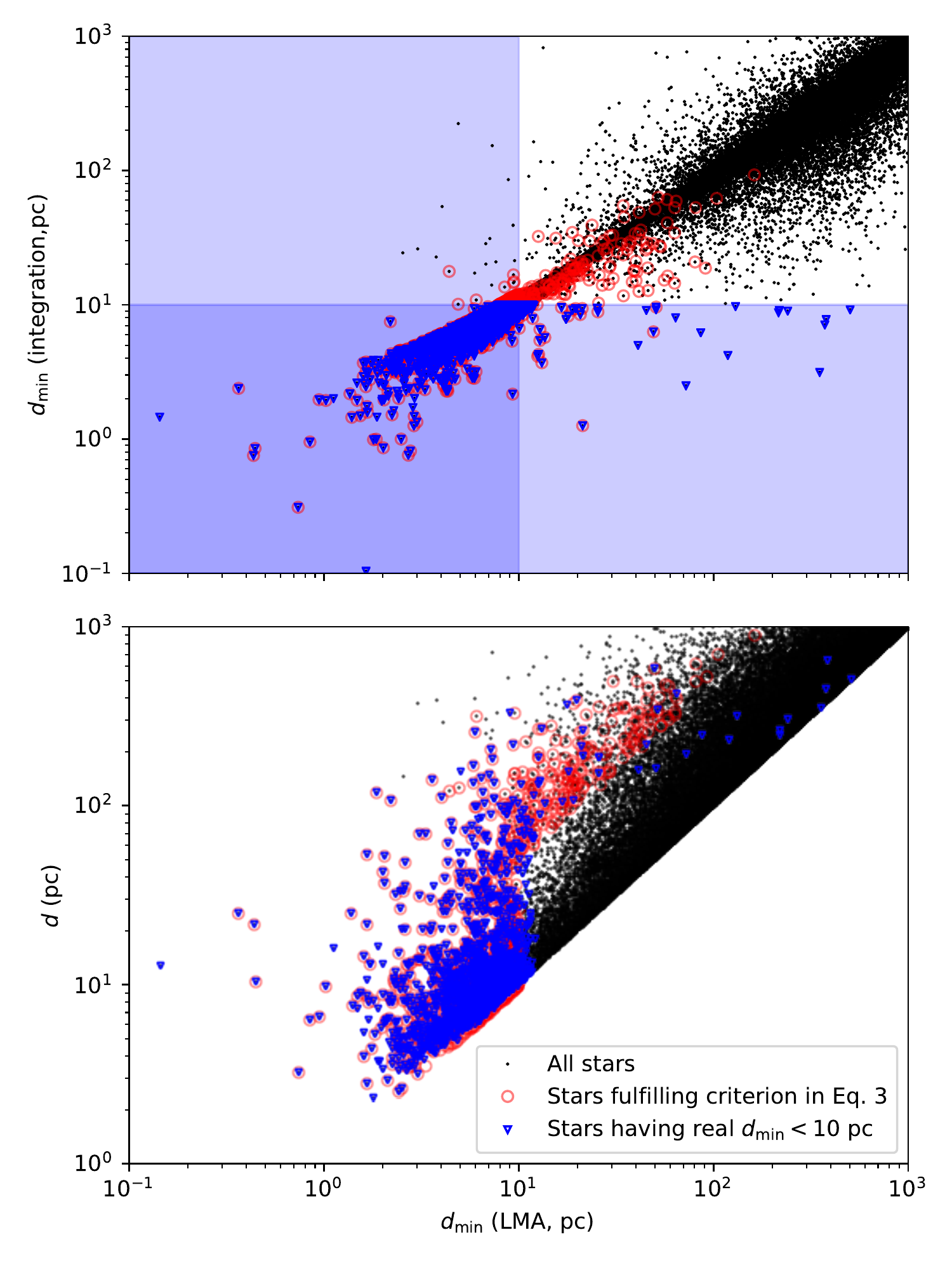}
\vspace{0.0cm}
\caption{\hl{Minimum encounter distances for the nominal orbit of \MUA\ and all the stars in the {\tt AstroRV} catalog (black dots), calculated with the LMA approximation (horizontal axis) and with a precise galactic potential integration (vertical axis of the upper panel).  Blue triangles represent the potential progenitors selected from their precise minimum distance, ie. $d\sub{min}<10$ pc; red circles mark the position of the stars fulfilling the heuristic selection criteria in \autoref{eq:CandidatesCriterion}.  The lower panel is intended to illustrate the way as this criterion was especially devised to minimize the number of actual potential progenitors that are missed when using only the LMA approximation (blue triangles without a corresponding red circle).} 
\label{fig:CandidateSelection}
}
}
\end{figure}

\hl{In the case of \MUA\ we have found that after applying the criterion in \autoref{eq:CandidatesCriterion}, only 2560 among the 285114 stars in the input catalog, were selected as candidates (see red circles in \autoref{fig:CandidateSelection}).  We verified with a full simulation (including only the nominal object and stars) that only a handful of suitable candidates (blue triangles in \autoref{fig:CandidateSelection}) were excluded from the probability calculations.}

\medskip 

For each ``candidate star'' selected with the preceding criterion, we should convert their observed properties (RA,DEC,$\varpi$,$\mu_\alpha$,$\mu_\delta$,$v_r$) (right ascension, declination, parallax, projected proper motion in RA, proper motion in declination and radial velocity, respectively) into its spatial cartesian coordinates in the galactic system, $(x\sub{gal},y\sub{gal},z\sub{gal},U,V,W)$ (and more importantly their corresponding uncertainties).  We use for this purpose the prescripcion \citealt{Johnson1987}. 

\subsection{Motion in the galactic potential}
\label{subsec:GalacticPotential}

\hl{As mentioned before,} the minimum LMA distances are useful at selecting the candidates but are a very crude approximation of the actual encounter conditions. The integrated effect of the galactic potential may modify substantially the position of the objects and the stars\hl{, especially after a long time of wandering in the interstellar space}.

In order to obtain a ``second order'' estimation of the minimum distances, we need to integrate the trajectory of the surrogate objects and the stellar candidates in the galactic potential.  Given the axisymmetric nature of the potential, we need to transform the cartesian coordinates of the objects with respect to the Sun ($x\sub{gal},y\sub{gal},z\sub{gal},U,V,W$) to cylindrical coordinates referred to the galactocentric reference system (see \autoref{fig:GalacticSystem}). This coordinate transformation is achieved in three steps:

\begin{enumerate}
\item Convert the position and velocities referred to the local standard of rest (LSR) into position and velocities relative to the galactic center:

$$
\vec{v}\sub{GC}=\vec{v}\sub{gal}+\vec{v}\sub{\odot}+v\sub{circ}\,\hat{y}\sub{gal}
$$

$$
\vec{r}\sub{GC}=\vec{r}\sub{gal}-R_\odot \hat{x}
$$

where $\vec{v}\sub{gal}:(U,V,W)$ is the velocity of the star with respect to the Sun in galactic coordinates, $\vec{v}\sub{\odot}:(U_\odot,V_\odot,W_\odot)$ is the velocity of the Sun with respect to the local standard of rest (LSR) and $v\sub{circ}$ is the local circular galactic velocity.  In \autoref{tab:GalacticParameters} we show the values adopted \hl{in this work} for these quantities.

\begin{figure*}
{
\centering
\includegraphics [width=150mm] {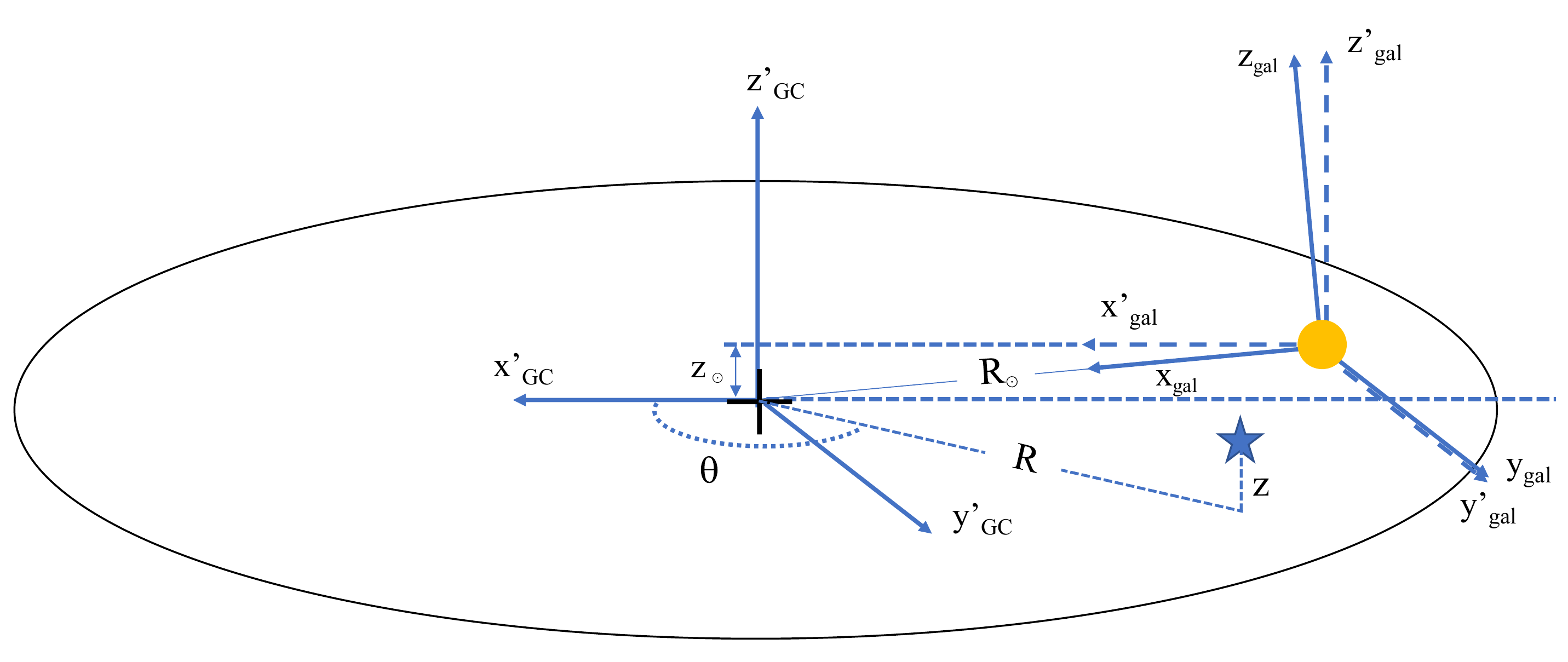}
\vspace{0.5cm}
\caption{Galactic Reference Systems.
\label{fig:GalacticSystem}
}
}
\end{figure*}

\item $\vec{v}\sub{GC}$ and $\vec{r}\sub{GC}$ are referred to a system pointing to the galactic center (the unprimed $x\sub{gal}$, $y\sub{gal}$, $z\sub{gal}$ in \autoref{fig:GalacticSystem}).  This system is rotated at an angle $\alpha=\sin^{-1}(z\sub{\odot}/R\sub{\odot})$ with respect to the plane of the Galaxy.  Thus, the actual physical galactocentric coordinates on which we must perform the orbital integration are obtained after the rotation:

\begin{eqnarray}
\nonumber
\vec{r'}\sub{GC} & = & R\sub{\alpha} \vec{r}\sub{GC}\\
\nonumber
\vec{v'}\sub{GC} & = & R\sub{\alpha} \vec{v}\sub{GC}
\end{eqnarray}

with 

$$
R\sub{\alpha}=
\left(
\begin{array}{ccc}
\cos\alpha & 0 & \sin\alpha\\
0 & 1 & 0\\
-\sin\alpha & 0 & \cos\alpha
\end{array}
\right)
$$

\item \hl{Finally, we need to} express the resulting galactocentric position and velocity in cylindrical coordinates\hl{, ie.} $\vec{r'}\sub{GC}$:($R$,$\theta$,$z$), $\vec{v'}\sub{GC}$:($\dot R$,$R \dot\theta$,$z$).

\end{enumerate}

In \hl{this coordinate system}, the equations of motion of an object moving in the potential of the galaxy $\Phi(R,\theta,z)$ are given by (see eg. \citealt{Garcia2001}):

\begin{eqnarray}
\label{eq:EoMGalactic}
\ddot R & = & -\frac{\partial\Phi}{\partial R} + R \dot\theta^2\\
\nonumber
\ddot \theta & = & -\frac{\partial\Phi}{\partial \theta} - 2 \frac{\dot R\dot \theta}{R} \\ 
\nonumber
\ddot z & = & -\frac{\partial\Phi}{\partial z}
\end{eqnarray}

Here we assume for simplicity an axisymmetric Kuzmin-like potential for the three galactic subsystems \citep{Kuzmin1956,Miyamoto1975}, disk (d), bulge (b) and halo (h):

\beq{label:GalacticPotential}
\Phi(R,\theta,z)=-\sum_{i=d,b,h}\frac{GM_i}{\sqrt{R^2+(a_i+\sqrt{z^2+b_i^2})^2}}
\eeq

The value of the potential parameters $M_i,a_i,b_i$ assumed for each component are summarized in \autoref{tab:GalacticParameters}

\begin{table}
  \centering
  \scriptsize
  \begin{tabular}{ccc}
  \hline Property & Value & Reference \\\hline\hline
  $U\sub{\odot}$ & 11.1 km/s & (1) \\
  $V\sub{\odot}$ & 12.24 km/s & (1) \\
  $W\sub{\odot}$ & 7.25 km/s & (1) \\
  $v\sub{circ}$ & 220.0 km/s & (2) \\
  $z\sub{\odot}$ & 17 pc & (3) \\
  $R\sub{0}$ & 8.2 kpc & (3) \\
  $M\sub{d},a\sub{d},b\sub{d}$ & $7.91\sci{10} M\sub{\odot}, 3500 pc, 250 pc$ & (4) \\
  $M\sub{b},a\sub{b},b\sub{b}$ & $1.40\sci{10} M\sub{\odot}, 0, 350 pc$ & (4) \\
  $M\sub{h},a\sub{h},b\sub{h}$ & $6.98\sci{11} M\sub{\odot}, 0, 24000 pc$ & (4) \\
  \hline
  \end{tabular}
\caption{Properties of the Galaxy.  References: (1) \citealt{Schonrich2010}, (2) \citealt{Bovy2015}, (3) \citealt{Karim2016}, (4) \citealt{Bailer2015}.
\label{tab:GalacticParameters}}
\end{table}

Once the trajectory of the objects in the potential of the Galaxy are integrated, we proceed at finding the encounter distance and time of the nominal object to each stellar candidate.  This is a refined estimation of $d\sub{min}$ and $t\sub{min}$.  


\section{The surrogate stars}

In the same way as we generate $N_p$ surrogate objects to take into account the uncertainties in the orbit solution of the interstellar interloper, we \hl{need to} generate, for each \hl{stellar candidate}, $N_s$ ``surrogate stars'' with observed properties compatible with the nominal astrometric observables, namely ($\alpha_0$,$\delta_0$,$\varpi_0$,$\mu\sub{\alpha,0}$,$\mu\sub{\delta,0}$).  For this purpose we build a covariance matrix (see \autoref{eq:Covariance}) using the errors reported for each astrometric variable and their related correlations\footnote{In the Hipparcos and Gaia databases, correlations are reported in the form of fields such as {\tt RA\_DEC\_CORR}, {\tt RA\_PARALLAX\_CORR}, etc.}.

\bigskip

\hl{Summarizing the previous sections,} studying the coincidence in time and space of an interstellar object with a nearby star, implies dealing properly with the kinematics of $N_p$ surrogate objects and $N_s$ surrogate stars. \hl{It is quite evident that} assessing the probability of a close encounter, using just the nominal solutions for the object and the stars is \hl{certainly} unrealistic. \hb{Moreover, dealing rigorously with the different reference system transformation and the galactic kinematics is neither optional.}

\section{Interstellar origin probability}
\label{sec:IOP}

\hb{We can now reformulate the origin probability question we raise in \autoref{sec:ProbabilityModel}, in terms of the properties defined in the previous section. The question is now:

\begin{quotation}
\noindent \em what is the probability that a star with present astrometric properties ($\alpha,\delta,\varpi,v_r$) ejected in the past a body that entered into the solar system with heliocentric orbital elements ($q,e,i,\Omega,\omega,t_p$) at a given reference epoch $t_0$.  
\end{quotation}
}

\begin{figure}
{
\centering
\includegraphics [width=90mm] {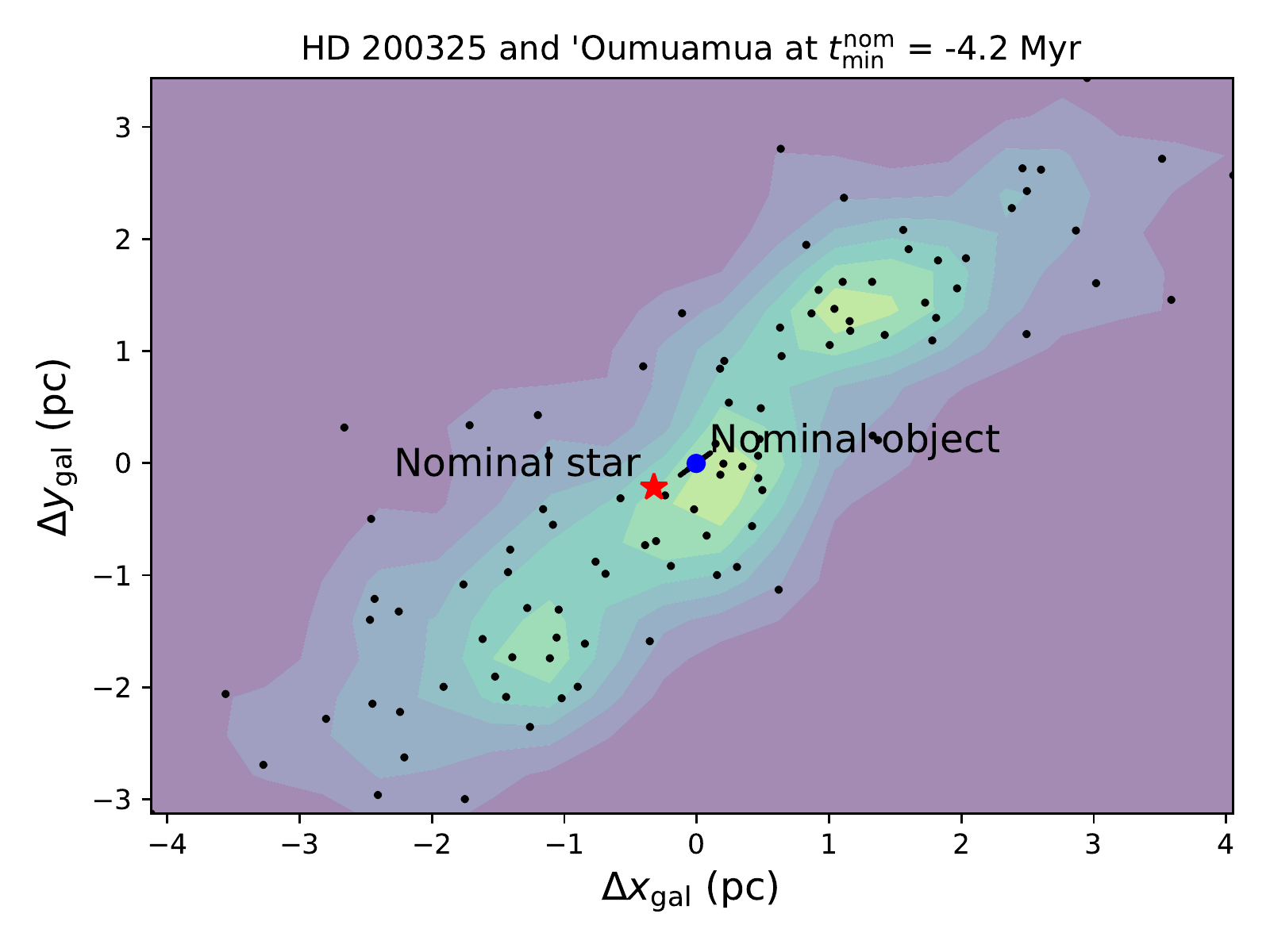}
\caption{Scattering plot of the position of the surrogate \hl{stars} corresponding to HD 200325 \hl{(red star marks its nominal position)} at the time of minimum distance with the nominal \hl{object} (blue dot). Contours show the number density of \hl{stars} \hl{around the object} estimated with the methods in this work. 
\label{fig:EncounterProbability}
}
}
\end{figure}

Let's assume first that the \hl{orbit of the interstellar object} is known with zero uncertainty.  If we propagate all the surrogate \hl{stars} until the time of minimum distance \hl{between the object and the nominal star}, $t\sup{nom}\sub{min}$, the \hl{object} will be surrounded by a cloud of points (see \autoref{fig:EncounterProbability}). Each of these points represents the position of \hl{one} surrogate \hl{star} at the time of encounter.  In the continuous limit, even if none of the $N_s$ surrogate \hl{stars} coincide in position with the \hl{object}, a probability different than zero exists that they were \hb{physically} connected.  Our central assumption here is that the probability that such a relationship actually existed will be proportional to the density of surrogate \hl{stars} at the position of the \hl{object}.  \hb{In terms of our probability model in \autoref{sec:ProbabilityModel}, the number $N\sub{pos}$ of parallel universes where the trajectory of the star and the object coincide will be proportional to the density of surrogate stars.}

Computing the number density from a discrete set of positions of the surrogate \hl{stars}, is challenging.  Several numerical techniques have been devised and applied in other areas such as cosmology and hydrodynamics (see eg. \citealt{Price2012}), to asses similar problems.  More recently, \citealt{Zuluaga2018} applied the approach used in Smooth Particle Hydrodynamics in the context of impact probabilities in the the Solar System.  According to this approach, the number density of \hl{stars} around \hl{an object with position $\vec r_p$} can be computed as:

\begin{eqnarray}
n(\vec r_p) &=& \sum_i^{N_s} W(|\vec r_p-\vec{r_*}_i|,h)
\label{eq:NumberDensity}
\end{eqnarray}
 
where $|\vec r_p-\vec{r_*}_i|$ is the distance between the \hl{object} and the $i$th surrogate \hl{star}, $h$ is a distance-scale for the distribution, and $W(d,h)$ is called a {\it smoothing kernel}. In this work we use for $h$, the characteristic size of the \hl{solar-mass truncation radius, namely $h\approx 0.5$ pc} \hb{(see \autoref{fig:ProbabilityModel})}.  Other prescriptions can be used, but for the purpose of testing our method we will restrict to this simple ansatz.

Different kernel function can be used for calculating $n$ \hl{as precisely as possible}.  Although it is common \hl{in SPH} to use a {\it B-spline kernel} (see eg. \citealt{Zuluaga2018}), for the purposes pursued here, the best suited function is one that provide a non-zero, although still very low value of $n(\vec r_p)$ for large values of $|\vec r_p-\vec{r_*}_i|$.  The kernel function \hl{used in this work will be a gaussian one} \citep{Price2012}:

\beq{eq:Wfunction}
W(d,h) = \sigma \exp(-d^2/h^2)
\eeq

where $\sigma=(\int W dV)^{-1}$ is a normalization constant.

\hl{We assume that} the probability that the candidate \hl{star} be at $t\sup{nom}\sub{min}$ inside a small volume $dV$ around the \hl{object} position will be given by:

\beq{eq:pdf}
p(\vec r_p)dV= \frac{n(\vec r_p) dV}{N_s}
\eeq

\hl{Since the number density in \autoref{eq:NumberDensity} has by definition the property $\int n\;dV = N_s$, the probability density function in \autoref{eq:pdf} is properly normalized.}

In the discrete case, if we take $\Delta V$ as \hb{the volume of a sphere having the truncation radius}, the probability of a coincidence in position between the star and the object at $t\sup{nom}\sub{min}$, $P\sub{pos}\sup{nom}$ can be estimated as:

\beq{eq:Ppos}
P\sub{pos}\sup{nom}\approx\frac{n(\vec r_p) \Delta V}{N_s}
\eeq

\hb{In the language of our probability model, there will be a number $N\sub{pos}\sup{nom}\propto P\sub{pos}\sup{nom}$ of parallel universes where the star and the object coincide in position.  However, as illustrated in \autoref{fig:ProbabilityModel}, encounter times depend on the region inside the intersection volume where the minimum approach happens. Therefore, the number $N\sub{pos}\sup{nom}$ will be just the number of encounters that are produced around that time.  We should integrate the cloud of surrogate objects and surrogate stars until a time $t\sub{min}\sup{i}$ where the closest approach between the object and the $i$-th star happens.  At that time we estimate the local number density of stellar trajectories, $n(\vec{r_p}^i)$ and the encounter probability $P\sub{pos}^i$:}

$$
P\sub{pos}^i\approx\frac{n(\vec{r_p}^i) \Delta V}{N_s}
$$

\hb{Finally, the total number of coincidences can be estimated by summing up the contributions $N\sub{pos}^i$, and the position probability can be finally written as:}

\beq{eq:Pposi}
P\sub{pos}\propto \frac{1}{N_s^2}\sum_i n(\vec{r_p}^i)
\eeq

\hb{Although the constant factor $N_s^2$ is common to all the stars in our simulations, we preserve it in order to have numbers of a reasonable order of magnitude.}


\subsection{Relative velocity}
\label{subsec:RelativeVelocity}

Only a few processes may lead to the ejection of a a small body from an almost-isolated planetary system \citep{Melosh2003,Napier2004}. In the low density solar neighborhood the more feasible ejection mechanisms are the particle-particle gravitational scattering, where small bodies receive a gravitational slingshot effect after encountering a planet \hl{or even a star in a multiple stellar system}, at the right conditions \citep{Raymond2017,Matija2018,Wiegert2014}. 
We estimate the distribution of excess velocities, $v\sub{\infty}$, that small bodies (asteroids and comets) receive from their encounters with a giant planet around a solar mass star, using a semi analytical approach inspired in the works by \citet{Wiegert2011,Wiegert2014}.  

For this purpose, we first set up a planetary system having a single planet of mass $M_p$ located in a circular orbit with semimajor axis $a_p$.  We randomly generate orbital elements for small bodies such that they intersect the orbit of the planet.  For simplicity the small-body semimajor axis, eccentricity and orbital inclinations were uniformly generated in the whole range of possible values, eg. $a\in(0.5 a_p - 1.5 a_p)$, $e\in(0,1)$, $i\in(0,90)$ deg.  Longitude of ascending node and argument of periapsis were calculated imposing the condition that the small body and the planet collide.

For each planet-small body orbit configuration we compute the relative velocity with which they encounter and the direction with respect to the planet reference frame from which the small body approaches. From here we follow the prescription of \citet{Wiegert2011} to compute the outbound velocity of the object after interacting with the planet.  A random position $(x_p,y_p)$ over the tangent plane to the Hill sphere is generated (see Figure 1 in \citealt{Wiegert2014}).  From there we compute the impact parameter, scattering angle, planetocentric orbital eccentricity and periapsis distance. Finally we rotate the planetocentric inbound velocity to compute the outbound velocity of the small-body at the Hill radius with respect to the planet and then with respect to the star.

Once the synthetic small bodies in our simulation are scattered by the planet, we evaluate if their outbound velocity $v$ with respect to the star is larger than the escape velocity $v\sub{esc}$ at the position of the planet. If this is the case, we compute the excess velocity, $v\sub{inf}^2=v^2-v\sub{esc}^2$.  

\begin{figure}
{
\centering
\includegraphics [width=80mm] {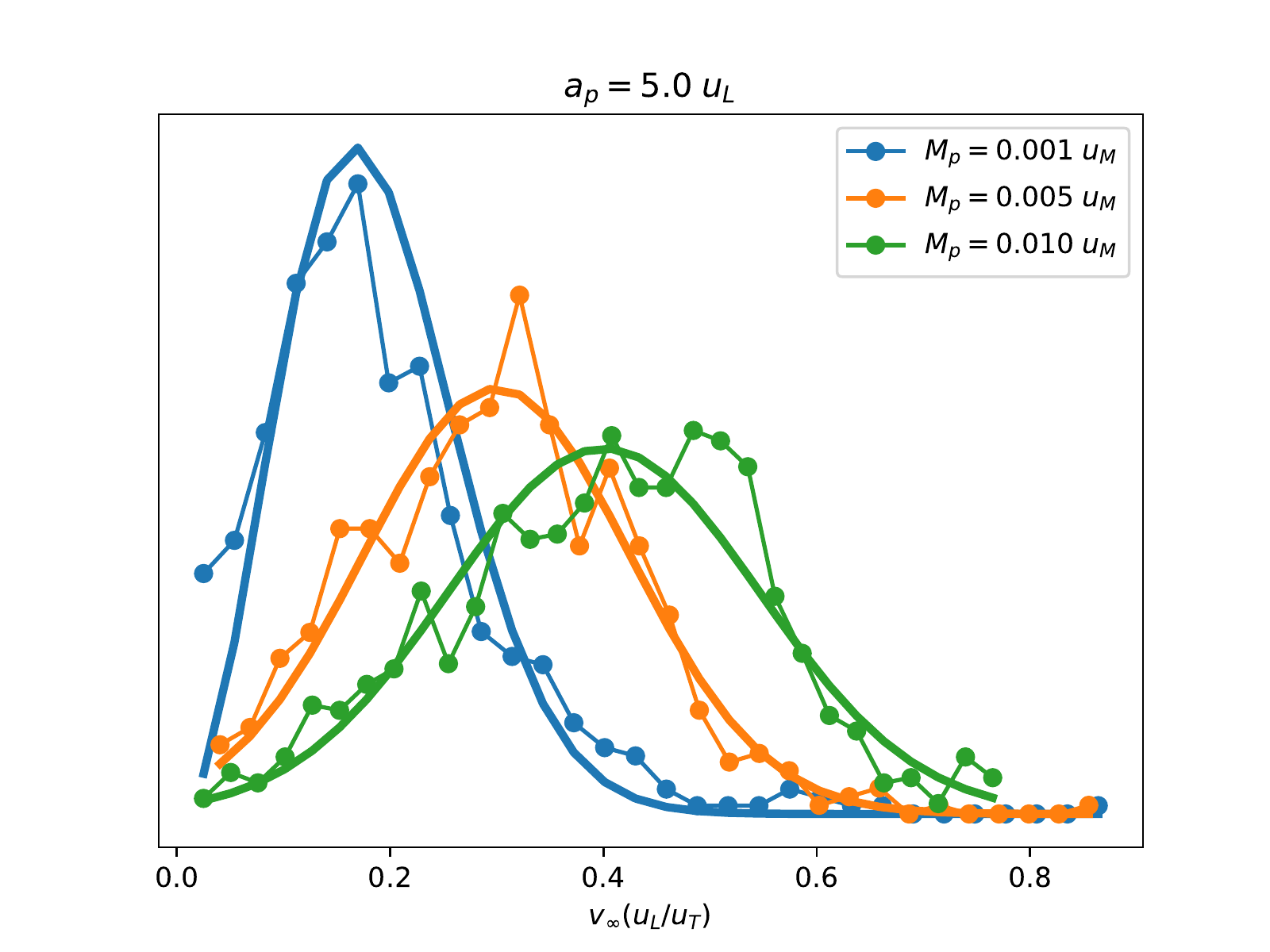}
\caption{Ejection velocity distribution for a planet in a circular orbit at $a_p=5\,u_L$ and different planetary masses.  Continuous thick lines are Maxwell-Boltzmann distributions with the same mean as the numerical results.
\label{fig:EjectionVelocity}
}
}
\end{figure}

In \autoref{fig:EjectionVelocity} we show the distribution of excess velocities resulting from the interactions of small bodies with planets of different mass $M_p$ located in a circular orbit $a_p=5$ AU around a solar mass-star.  The results are given in canonic units for which we have set $G=1$, $u_L=1$ AU and $u_M=1\,M\sub{\odot}$.  In these units, $u_T=\sqrt{u_L^3/(G u_M)}$, and $u_L/u_T$= 30 km/s. 

An interesting advantage of expressing the results in canonic units is that they can be used to predict ejection velocities distribution from planetary systems around stars of arbitrary mass.  Thus, for instance, the average \hl{ejection} velocity for a planet with the mass of Jupiter around a solar mass star, $u_M=1\;M_\odot$, $M_p=0.001\;u_M$ is $\bar{v}\sub{\infty}=0.2\;u_L/u_T$, or 6 km/s (see the leftmost curve in \autoref{fig:EjectionVelocity}).  If we consider now an \hl{early} M-dwarf with mass $M\sub{\star}=0.5\;M\sub{\odot}$, the same results in \autoref{fig:EjectionVelocity} will apply, but now for the case of a planet with half the mass of Jupiter. The value of $\bar{v}\sub{\infty}$ in km/s, will depend on what value is assumed for $u_L$.  If we take $u_L=1$ AU (as in the case of the solar-mass star), then $u_L/u_T=20$ km/s for the \hl{early} M-dwarf, and the average ejection velocity will be 4 km/s.

\begin{figure}
{
\centering
\includegraphics [width=90mm] {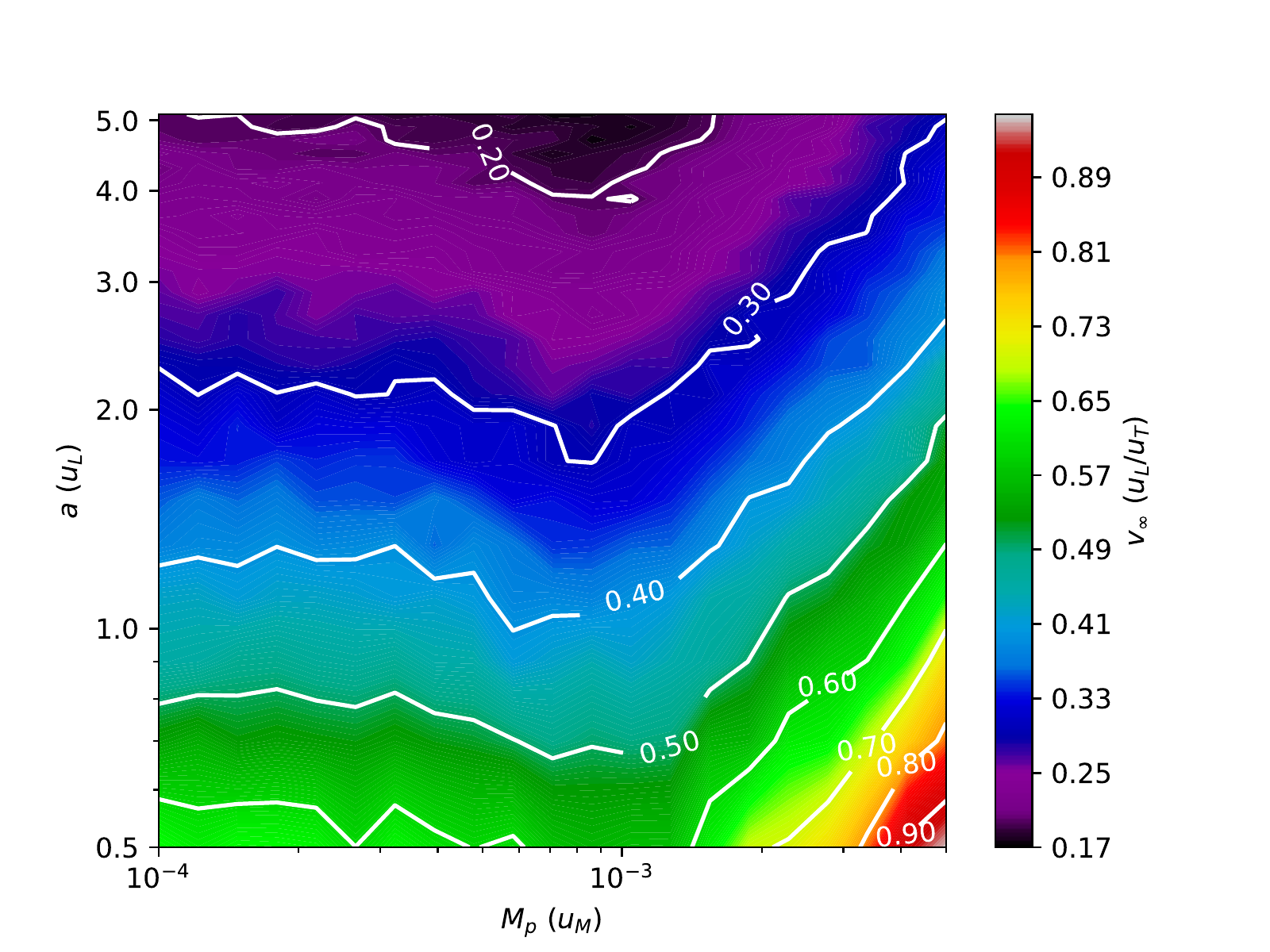}
\includegraphics [width=80mm] {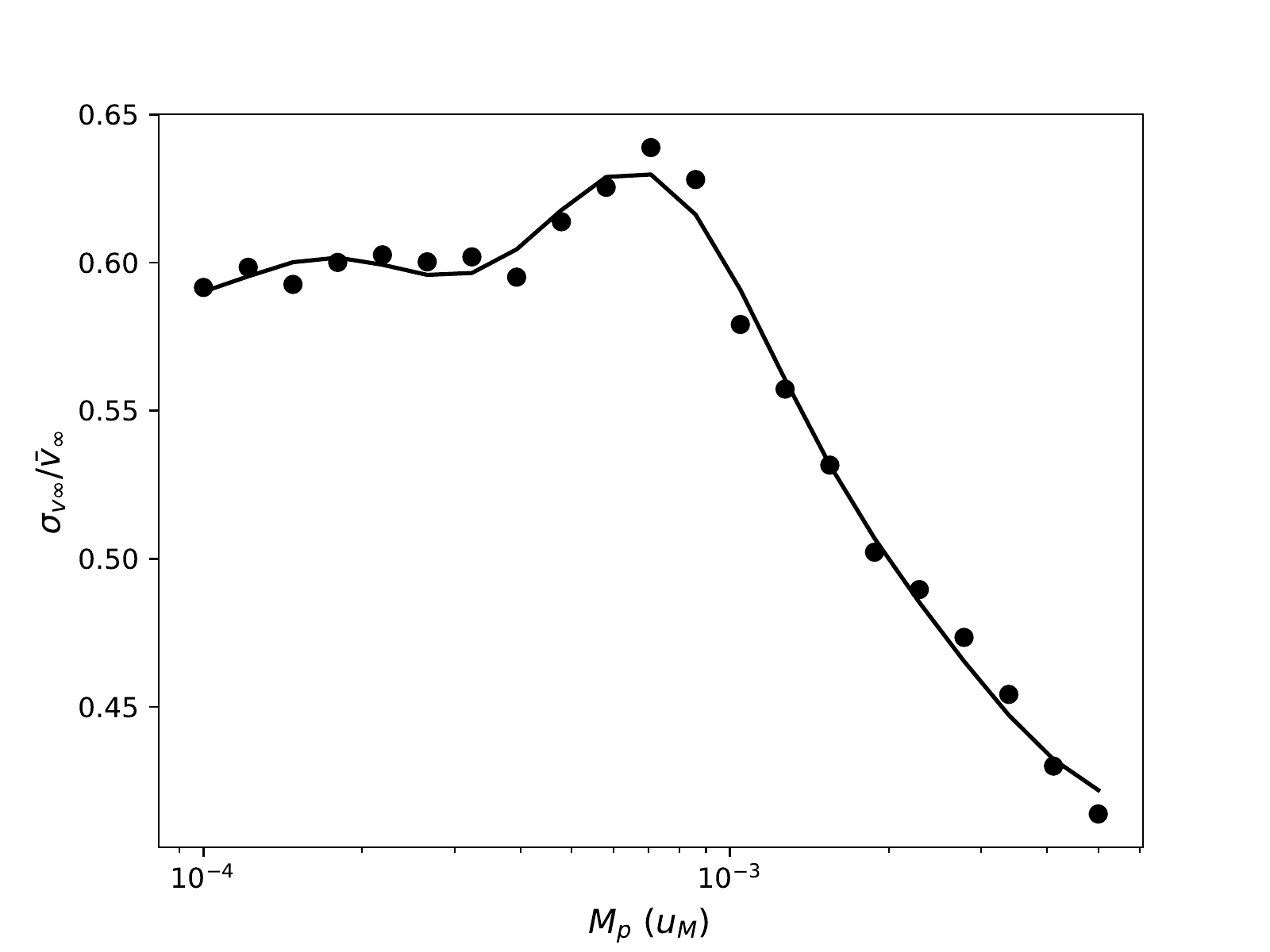}
\caption{Upper panel: ejection mean velocity for different planetary masses and orbital sizes. Lower panel: ratio of the average to the standard deviation of the ejection velocities for different planetary masses. 
Properties are given in canonic unites.  If $u_L=$1 AU and $u_M=1\,M\sub{\odot}$, $u_v=u_L/u_T=30$ km/s.
\label{fig:EjectionVelocityContours}
}
}
\end{figure}

In \autoref{fig:EjectionVelocityContours} we show contours of $\bar{v}\sub{\infty}$ in the $a_p-M_p$ plane.  We discover that for masses below $10^{-3}\;u_M$ ejection velocities are only a function of planetary orbital distance and are very insensitive to planetary mass.

Another interesting result from our semi analytical experiments is that the ratio of the standard deviation $\sigma_{v\infty}$ to the mean value of the ejection velocity $\bar{v}\sub{\infty}$, is almost independent of planetary orbital distance.  For its dependency on planetary mass, in the lower panel of \autoref{fig:EjectionVelocityContours} we show a plot of the value of $\sigma_{v\infty}/\bar{v}\sub{\infty}$ for different planetary masses. It is interesting to notice that in the case of a Maxwell-Boltzmann distribution (MBD) the ratio $\sigma/\mu$ (with $\mu$ the mean) is constant and equal to 0.42 which is of the order of $\sigma_{v\infty}/\bar{v}\sub{\infty}$ \hl{in our own experiments}.  This seems to suggest that the ejection velocities can be fitted by a MBD with a mean that depends on $M_p$ and $a_p$.  This is precisely the fitting functions we have used in \autoref{fig:EjectionVelocity}.

\begin{figure}
{
\centering
\includegraphics [width=90mm] {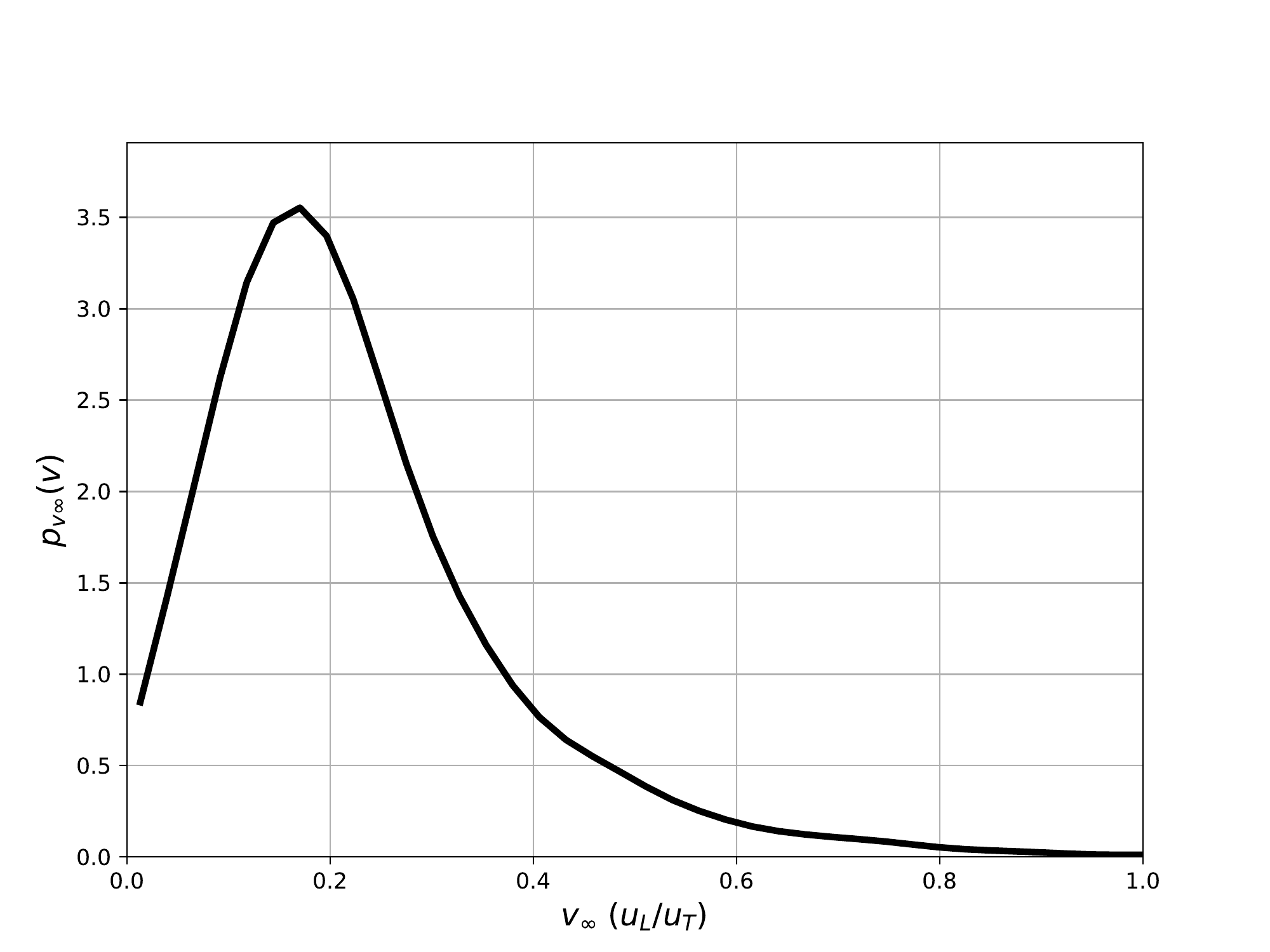}
\caption{Ejection velocity posterior distribution as estimated in this work.
\label{fig:EjectionVelocityPosterior}
}
}
\end{figure}

Using the average ejection velocities and the dispersion-to-mean ratio in \autoref{fig:EjectionVelocityContours}, we can estimate the velocity distribution of small-bodies being ejected from a planetary system around a star of a given mass.  Since ejection velocities depend on the unknown mass $M_p$ and semimajor axis $a_p$ of the largest planet in the system, we estimate the ``posterior'' ejection velocity probabilities, assuming for simplicity uniform ``priors'' for these quantities.  The resulting posterior distribution $p_{v\infty}(v)$ for the case of a solar-mass star is shown in \autoref{fig:EjectionVelocityPosterior}.

\bigskip

\hb{Once we have a posterior probability distribution for ejection speeds (see \autoref{fig:EjectionVelocityPosterior}), the factor $f\sub{ind}$ in the original IOP expression (\autoref{eq:IOP}) can be estimated as:}

\beq{eq:find}
f\sub{ind}^i \propto \left\{
\begin{array}{cc}
p(\vrel^i) & p_B\gg p\sub{ind} \\
1 & p\sub{ind}\gg p_B \\
\end{array}
\right.
\eeq

\hb{``integrating'' across the intersection volume and assuming that the number of background objects coming out from the stellar system is much larger than the number of indigenous objects, the joint position-velocity probability $P\sub{pos,vel}$ will be proportional to:}

\beq{eq:Pposvel}
P\sub{pos,vel} \propto 
\frac{1}{N_s^2} \sum_i n(\vec{r_p}^i) p(\vrel^i) 
\eeq

\hb{Finally the IOP for the star will be}:

\beq{eq:FinalIOP}
{\rm IOP}=
\left\{
\begin{array}{cc}
P\sub{pos} & p\sub{ind}\gg p_B \\
P\sub{pos,vel} & p_B\gg p\sub{ind} \\
\end{array}
\right.
\eeq

\subsection{Normalization of the IOP}
\label{subsec:Normalization}

\hb{

The normalization of the interstellar origin probability will depend on what we define as the ``sample universe'' $\Omega$ of our probability space. 

If we assume for instance, that all interstellar objects coming into the Solar System are ejected from stars in the Galaxy, the sample universe will be $\Omega=\bigcup_n s_n$, where $s_n$ is the event ``the interstellar object comes from the $n$-th star''.  If we assume that all the event $s_n$ are independent, then:

$$
P(\Omega)=\sum_n \mathrm{IOP}\sup{(n)}=1
$$

where IOP$\sup{(n)}$ is the interstellar origin probability of the $n$-th star as computed with \autoref{eq:FinalIOP}.  If this case, the normalization constant, following \autoref{eq:IOP}, will simply be:

\beq{eq:Normalization}
{\cal N}\sub{stellar}=\left(\sum_n \mathrm{IOP}\sup{(n)}\right)^{-1}
\eeq

and the normalized IOP probability of the $n$-th star can be obtained multiplying the value in the right hand side of \autoref{eq:Ppos} or \autoref{eq:Pposvel} by this constant.

If on the other hand, we admit that some interstellar objects could come from processes different than the ejection from a stellar system, then the sample space will be larger, and hence ${\cal N}\sub{stellar}$ will be an overestimation of the actual normalization constant.

Moreover, since we have in our experiments a limited set of stars (nearby stars with complete astrometric information), the normalization constant estimated with \autoref{eq:Normalization} will also represent an overestimation of the true one.  Therefore the ``normalized'' IOP probability will also overestimate the true one.  

Still, the IOP computed under our assumptions and with a sample-limited normalization, will be good enough to sort-out our potential progenitors and to concentrate our potential follow-up efforts in those having the largest IOP values.

}

\section{Results}
\label{sec:Results}

As an illustrative example (not necessarily the best one, but the only to date), we apply our methodology to assess the interstellar origin of \MUA.  In \autoref{tab:Progenitors} we present a list of the potential progenitors satisfying \hl{the conditions} $t\sup{nom}\sub{min}<t\sub{Ret}\approx 40$ Myr and $d\sup{nom}\sub{min}<2$ pc.  \hl{For each progenitor we present several statistics of the encounter conditions, namely, encounter time $t\sub{min}$, minimum encounter distance $d\sub{min}$, and relative velocity $v\sub{rel}$.  Along with the nominal value of these quantities (first row of each entry) we provide the value of the 10\%, 50\% (median) and 90\% percentiles.} \hll{In the case of  minimum distance, providing the value of the percentiles is uninformative if the cloud of points representing the relative position of surrogate objects and surrogate stars, surrounds the nominal relative position.  In this case we have calculated and reported a new statistics, the ``centering parameter'' $f$, defined as the fraction of points at a distance less than or equal to $d\sub{min,nom}$ from the nominal relative position. If the cloud of objects is perfectly centered around the nominal relative position (an ideal configuration for an actual progenitor), $f=0$. If on the other hand the cloud is off by several times its own dispersion, then $f\sim 0.5$ (independent of  distance).  When the cloud is centered, it is better to provide the minimum value of $d\sub{min}$ which is precisely the second number between brackets in Table 4.  When the cloud is decentered, it is better to read the 10\% and 90\% percentiles which are provided as the third and forth figures between brackets.}

\hl{In the last two columns of the table the value of the IOP are reported.} \hb{We have included both, the value of $P\sub{pos}$ and $P\sub{pos,vel}$.  Thus, the IOP can be judged according to the the two extreme cases in \autoref{eq:FinalIOP}.  In all cases, the IOP values have been normalized following the procedure describe in \autoref{subsec:Normalization}.}

To provide an idea of the astrometric uncertainties involved in the calculation of the origin probability, we have tabulated for each potential progenitor, an ``astrometric quality index'' $q$, defined as the minimum ratio between the value of each astrometric key property (parallax, proper motion and radial velocity) and its standard error.  Therefore, a quality factor of 1 implies that one of these quantities has an error of the same order than its magnitude (poor astrometry).  On the other hand, a large $q$-value are indicative of the availability of very precise astrometric properties (including radial velocity). 

\begin{table*}
\centering
\scriptsize
\begin{tabular}{llll|ccc|ccc}
\hline
\multicolumn{4}{c|}{Basic properties} & 
\multicolumn{3}{c|}{Encounter conditions} & \multicolumn{3}{c}{$\log\;$IOP}  \\ \hline
\# & Name & $d_*$ (pc) & $q$ & 
$t\sub{min}$ (Myr)  & 
$d\sub{min}$ (pc)   & 
$v\sub{rel}$ (km/s) & 
$P\sub{pos}$ & $\langle f_{\rm ind}\rangle$ & $P\sub{pos,vel}$ \\
\hline\hline
 1 & HIP 103749 & 53.8 & 1 & ${-4.22}$ &${1.75}$ &${12.0}$ &${-1.6}$ & ${-1.9}$ & ${-0.5}$ \\
 & (\href{http://simbad.u-strasbg.fr/simbad/sim-id?Ident=HD%20200325}{HD 200325}) & & & \tiny $[-4.41,-4.22,-4.05]$ &\tiny $[f=0.55,{\rm min}=0.31,0.37,4.98]$ &\tiny $[11.4,12.0,12.5]$ & & & \\\hline

 2 & TYC 3144-2040-1 & 4.5 & 2 & ${-0.12}$ &${1.00}$ &${17.9}$ &${-1.8}$ & ${-2.9}$ & ${-1.6}$ \\
 &  & & & \tiny $[-0.12,-0.11,-0.11]$ &\tiny $[f=0.60,{\rm min}=0.86,0.88,1.14]$ &\tiny $[17.3,18.0,18.5]$ & & & \\\hline

 3 & TYC 7069-1289-1 & 8.3 & 1 & ${-0.39}$ &${0.99}$ &${24.6}$ &${-1.8}$ & ${-3.7}$ & ${-2.3}$ \\
 &  & & & \tiny $[-0.41,-0.38,-0.26]$ &\tiny $[f=0.30,{\rm min}=0.53,0.61,5.25]$ &\tiny $[23.3,25.1,30.6]$ & & & \\\hline

 4 & HIP 3821 & 6.0 & 85 & ${-0.17}$ &${1.26}$ &${23.5}$ &${-2.8}$ & ${-3.4}$ & ${-3.1}$ \\
 & (\href{http://simbad.u-strasbg.fr/simbad/sim-id?Ident=*%20eta%20Cas}{* eta Cas}) & & & \tiny $[-0.17,-0.17,-0.17]$ &\tiny $[f=0.60,{\rm min}=1.23,1.23,1.27]$ &\tiny $[23.3,23.5,23.6]$ & & & \\\hline

 5 & TYC 3663-2669-1 & 6.1 & 34 & ${-0.17}$ &${1.34}$ &${23.9}$ &${-3.0}$ & ${-3.5}$ & ${-3.4}$ \\
 &  & & & \tiny $[-0.17,-0.17,-0.16]$ &\tiny $[f=0.65,{\rm min}=1.13,1.20,1.46]$ &\tiny $[23.1,23.8,24.3]$ & & & \\\hline

 6 & HIP 91768 & 3.5 & 11 & ${-0.03}$ &${0.82}$ &${36.8}$ &${-1.2}$ & ${-5.5}$ & ${-3.6}$ \\
 & (\href{http://simbad.u-strasbg.fr/simbad/sim-id?Ident=HD%20173739}{HD 173739}) & & & \tiny $[-0.03,-0.03,-0.03]$ &\tiny $[f=0.50,{\rm min}=0.81,0.81,0.82]$ &\tiny $[36.7,36.8,36.9]$ & & & \\\hline

 7 & HIP 91772 & 3.5 & 10 & ${-0.03}$ &${0.76}$ &${39.3}$ &${-1.1}$ & ${-5.8}$ & ${-3.8}$ \\
 & (\href{http://simbad.u-strasbg.fr/simbad/sim-id?Ident=HD%20173740}{HD 173740}) & & & \tiny $[-0.03,-0.03,-0.03]$ &\tiny $[f=0.55,{\rm min}=0.75,0.75,0.76]$ &\tiny $[39.1,39.3,39.4]$ & & & \\\hline

 8 & TYC 6573-3979-1 & 6.5 & 2 & ${-0.18}$ &${0.95}$ &${44.6}$ &${-0.8}$ & ${-6.6}$ & ${-4.3}$ \\
 &  & & & \tiny $[-0.18,-0.18,-0.17]$ &\tiny $[f=0.55,{\rm min}=0.24,0.27,1.81]$ &\tiny $[44.1,44.7,45.8]$ & & & \\\hline

 9 & HIP 18453 & 37.4 & 30 & ${-0.86}$ &${0.86}$ &${41.0}$ &${-1.8}$ & ${-6.1}$ & ${-4.8}$ \\
 & (\href{http://simbad.u-strasbg.fr/simbad/sim-id?Ident=*%2043%20Per}{* 43 Per}) & & & \tiny $[-0.87,-0.86,-0.85]$ &\tiny $[f=0.25,{\rm min}=0.76,0.80,1.42]$ &\tiny $[40.6,41.1,41.5]$ & & & \\\hline

 10 & TYC 7582-1449-1 & 192.2 & 1 & ${-8.96}$ &${1.26}$ &${22.1}$ &${-4.6}$ & ${-3.6}$ & ${-5.3}$ \\
 &  & & & \tiny $[-9.20,-8.83,-8.59]$ &\tiny $[f=0.05,{\rm min}=1.26,2.59,25.59]$ &\tiny $[21.8,22.4,23.3]$ & & & \\\hline

 11 & TYC 7142-1661-1 & 21.7 & 1 & ${-0.62}$ &${0.75}$ &${36.9}$ &${-2.9}$ & ${-5.7}$ & ${-5.4}$ \\
 &  & & & \tiny $[-0.62,-0.59,-0.54]$ &\tiny $[f=0.06,{\rm min}=0.75,1.43,14.50]$ &\tiny $[36.8,37.9,47.4]$ & & & \\\hline

 12 & HIP 63797 & 118.1 & 11 & ${-2.90}$ &${1.00}$ &${40.2}$ &${-2.5}$ & ${-5.7}$ & ${-5.8}$ \\
 & (\href{http://simbad.u-strasbg.fr/simbad/sim-id?Ident=HD%20113376}{HD 113376}) & & & \tiny $[-3.33,-2.83,-2.49]$ &\tiny $[f=0.25,{\rm min}=0.50,0.67,3.28]$ &\tiny $[35.0,41.2,46.8]$ & & & \\\hline

 13 & HIP 101180 & 8.1 & 52 & ${-0.17}$ &${1.58}$ &${32.6}$ &${-4.4}$ & ${-4.9}$ & ${-6.2}$ \\
 & (\href{http://simbad.u-strasbg.fr/simbad/sim-id?Ident=LHS%203558}{LHS 3558}) & & & \tiny $[-0.17,-0.17,-0.16]$ &\tiny $[f=0.50,{\rm min}=1.57,1.58,1.60]$ &\tiny $[32.4,32.6,32.7]$ & & & \\\hline

 14 & TYC 7093-310-1 & 6.7 & 1 & ${-0.19}$ &${1.96}$ &${40.3}$ &${-3.7}$ & ${-5.9}$ & ${-6.5}$ \\
 &  & & & \tiny $[-0.21,-0.20,-0.19]$ &\tiny $[f=0.50,{\rm min}=1.08,1.56,2.50]$ &\tiny $[38.6,40.2,41.7]$ & & & \\\hline

 15 & HIP 1475 & 3.6 & 106 & ${-0.03}$ &${1.47}$ &${38.7}$ &${-3.8}$ & ${-5.8}$ & ${-6.5}$ \\
 & (\href{http://simbad.u-strasbg.fr/simbad/sim-id?Ident=V*%20GX%20And}{V* GX And}) & & & \tiny $[-0.03,-0.03,-0.03]$ &\tiny $[f=0.65,{\rm min}=1.47,1.47,1.47]$ &\tiny $[38.6,38.7,38.8]$ & & & \\\hline

 16 & HIP 21553 & 9.9 & 171 & ${-0.24}$ &${1.94}$ &${34.8}$ &${-6.6}$ & ${-5.2}$ & ${-8.7}$ \\
 & (\href{http://simbad.u-strasbg.fr/simbad/sim-id?Ident=HD%20232979}{HD 232979}) & & & \tiny $[-0.25,-0.24,-0.24]$ &\tiny $[f=0.45,{\rm min}=1.90,1.91,1.96]$ &\tiny $[34.6,34.8,34.9]$ & & & \\\hline

\hline
  \end{tabular}
\caption{Interstellar origin probability (IOP) for a selected group of nearby stars.
\label{tab:Progenitors}}
\end{table*}

\hl{Using the available information, we identify only 16 potential progenitors for \MUA\ fulfilling all the selection criteria.  Most of the potential progenitors have moderately large $q$-values and are located at distances beyond 5 pc.  As expected the IOP probability has achieved at selecting candidates with moderate relative velocities and encounter distances between 0.2 and 5 pc (with a few exceptions, eg. TYC 7582-1449-1 that also has a poor astrometry).} 

The method presented here does not necessarily intend to identify a single object as the actual progenitor of \MUA.  Finding the origin would require follow-up observations of the interstellar object (while reachable), and improving the astrometric properties of the \hl{potential progenitors}.  When more and better information be available about these and other stars, the list could be extended or reduced, and more importantly the IOP probability could be modified. 

Still, it is interesting to notice the properties of several of these \hl{progenitor} candidates.  

The most interesting case is of course that of HD 200325 (HIP 103749), the first object in the list.  The star is probably a double or multiple system \citep{Cvetkovic2011}.  It is located at the present at a distance of $53.8$ pc from the Sun and their physical properties are well constrained (see eg. \citealt{Cvetkovic2011,Holmberg2007}).  Its radial velocity has been measured very precisely ($v_r=-11.10\pm 0.4$ km/s) and their astrometric properties are \hl{relatively well known (the object is in the Hipparcos catalog but not in the GAIA TGAS set)}.  Its most uncertain astrometric property is the declination proper motion $\mu\sub{dec}=0.90\pm 0.60$ mas, \hl{which is consistent with its low $q$-value}. We expect that \hb{better astrometric parameters be obtained and published in the next Gaia Data Release.}  \hl{The uncertain proper motion is the} reason why the minimum encounter distance has a large uncertainty, ie. \hl{0.5-5} pc.  The mass of the main component of the HD 200325 system is $1.19\pm 0.1\;M\odot$, and its age is around $3.2$ Gyr. The companion seems to be a low mass K-dwarf \citep{Cvetkovic2011} located at $\sim$ 25 AU from the primary. Although no planet has been discovered yet around the primary star, and its binary nature may prevent the formation of giant planets \citep{Thebault2011}, the stars are far apart and their masses are very dissimilar.  Interestingly, there is a known binary system with similar characteristics, HD 41004 (a solar mass evolved primary with a low-mass companion at $~$ 20 AU) around which a jupiter-mass planet has been discovered at $\sim$ 1 AU from the primary \citep{Zucker2004}.  The existence of this ``doppelg\"anger'', together with recent theoretical evidence that shows that formation of planets around this kind of binaries could not be as improbable as thought \citep{Higuchi2017}, lead us to speculate that HD200325 may harbor a planetary system and probably be the source of ejected small bodies. \hl{The possibility that \MUA\ be an ejecta of a binary system has been already considered by other authors \citep{Matija2018,Raymond2017} which give some theoretical support to our speculation.}

\hl{Our results match well the works by \citet{Matija2018} and \citet{Raymond2017}, that predict a non-catastrophic origin of \Oumuamua\ in the neighborhood of a binary stellar system.}  

It is, however, too early to conclude that \MUA\  comes actually from our best \hl{potential progenitor}.  \hl{It was not either our aim proposing it}.  \hl{However,} We expect that improved astrometric information about \hl{this and other stellar systems included in our {\tt AstroRV} catalogue, be available} with the Gaia Data Release 2 and help us to improve the IOP for our best candidates \hl{or to find even better ones}.  

\section{Discussion}
\label{sec:Discussion}

When dealing with very uncertain processes such as those involved in this problem, it is important to ask if the identified close encounters could be just the product of chance.  Further numerical experiments should be performed to test this idea and will be presented in a future work.  

At least three groups, that of \citet{Portegies2017},  \citet{Dybczyski2017} and \citet{Feng2018} published their own list of candidates.  Some of their objects \hl{are among the candidates in the list in \autoref{tab:Progenitors}}, but others are not there.  We searched for the \hl{``missing''} objects among our {\tt AstroRV} catalog and find that either some of their candidates were not included in our input catalog or have properties (relative velocities, time for minimum approach) too large for our particular selection criteria.  This fact put in evidence a limitation of any approach to asses the origin of an interstellar object: the completeness of the database. 

The approach presented here to estimate ejection velocities of small bodies from planetary system, is only our first attempt to model what \hb{should be for sure} a more complex problem.  Although a lot of interest have been paid in to model the flux of planetesimals coming out from young planetary systems, predicting the direction and velocities of these ejected objects has received less attention.  The case of interstellar objects and the investigation of their origin could encourage more research in this field.  Thus, for instance, improved semi-analytical models and detailed numerical n-body simulations may be required to better constraint the kinematical properties of ejected small bodies from already formed planetary systems around single and multiple stellar systems. We have already performed several basic n-body simulations to investigate the problem that confirms some of our semi analytical results but also seems to predict lower ejection velocities in some regions of the parameter space.

Although trillions of interstellar small objects are wandering around the Solar System and most of them could be there for hundreds of millions if not billions of years, the effort for tracing back the origin of some of those that enter for chance into the inner Solar System, is not irrelevant.  Although many stars may have contributed in the history of Galaxy to populate this graveyard, of course nearby stellar system could be an important source of many of these objects.

Assessing the origin of interstellar objects require that the small uncertainties in the initial kinematic parameters do not propagate into large errors in the resulting dynamical properties due to factors related with the simulation process.  Some sources of errors include but are not restricted to galactic coordinate transformation, uncertainties in the galactic parameters and of course errors in coding and processing the data. 

In the same way as the trajectory can be propagated backward, it could also be propagated forward in time \hl{to predict the fate of these interstellar interlopers.  At studying their fate we can also learn interesting things that could shade some light on their own origin.}

\section{Summary and conclusions}
\label{sec:SummaryConclusions}

In this paper we presented a general method for calculating the probability that nearby stars be the source of an interstellar small object detected inside the Solar System.  The method relies on the availability of a precisely determined orbit for the object and precise astrometric information about a large enough number of nearby stars.  For illustrating the method we applied it for assessing the origin of \MUA, the first \hl{identified} interstellar interloper.

The application of our method to the case of \MUA, allowed us to identify a handful of stars whose kinematical and physical properties are compatible with the ejection of a small object in the latest couple of Myrs.  Of particular interest, at least with the available information, is the binary(multiple) system HD200325.  The system is dominated by a primary star $1.2\;M\sub{\odot}$ with a K-dwarf companion at $\sim$ 25 AU.  Although no planetary system have been discovered yet around the primary or the secondary star, several similar multiple systems have been discovered in the past with planets; this fact, together with multiple recent theoretical evidences, suggest that the case for HD200325 as \MUA\  progenitor is not as unlikely as previously thought.

Our method is not intended to identify a single object as the definitive progenitor.  Even with small uncertainties in the initial orbit and in the astrometric parameters of the stars, there will be always large enough uncertainties in the resulting kinematical properties that constrained our capability to pinpoint a single source.  Our aim is to identify stars whose properties could be studied in more detail to reduce the uncertainties and increase/decrease the probability that they can be the sources of these objects.

One of the most interesting \hl{features} of our method is \hl{the fact} that IOP probabilities can be published and updated permanently when new and better information be obtained.  A catalog of \hl{potential progenitors} for this and other future discovered interstellar objects can also be compiled and published with a global ranking of IOP probabilities.  The authors believe that it could be a time in the future when this and other efforts could allows us to pinpoint precisely the provenance of an interstellar object.  Those will be the times when instead of going to other planetary systems we will be able to study them using natural probes flying through our Solar System.

\section*{Acknowledgements}

We thank Coryn Bailer-Jones for providing some of the data used in this work to compile the AstroRV catalog. We appreciate all the observations and suggestions received from our fellow colleagues F. Feng, E. Mamajek, M. {\'C}uk and S. Raymond.  We express our gratitude to the anonymous referee that carefully revised the manuscript and provide insightful comments and corrections that make possible the final version of this work. Some of the computations that made possible this work were performed with {\tt Python 3.5} and their related tools and libraries, {\tt iPython} \citep{Perez2007}, {\tt Matplotlib} \citep{Hunter2007}, {\tt scipy} and {\tt numpy} \citep{Van2011}. This work has made use of data from the European Space Agency (ESA) mission {\it Gaia} (\url{https://www.cosmos.esa.int/gaia}), processed by the {\it Gaia} Data Processing and Analysis Consortium (DPAC, \url{https://www.cosmos.esa.int/web/gaia/dpac/consortium}).

\appendix

\section{{\tt iWander} package}

We have provided with this work an open source package, {\tt iWander}, that implements the general methodology described here.  Providing a fully fledged computational tool is not only an effort to make these results reproducible, but also to allow the methodology to be applied by any researcher once future interstellar objects be discovered.  The package can also serve as a basis or an inspiration to develop better computational tools for this and other related problems.

Here we provide some basic information about the package that can be useful for users and developers:

\begin{itemize}

\item The package is available at {\tt GitHub}: \url{http://github.com/seap-udea/iWander}.

\item The required {\tt NASA NAIF SPICE} kernels as well as the libraries required to compile them, are provided with the package while complying the corresponding licenses, in order to ease its compilation and use.

\item The core modules of the package were written in {\tt C/C++} to guarantee computing efficiency.  As a result they should be compiled before usage.  Other post processing modules were written in {\tt Python} and are provided as core python scripts as well as {\tt iPython notebooks}.

\item One of the key components of the methodology and the package are the astrometric and radial velocities catalogs.  Although all of them are publicly available, we also provide them with the package.  \hl{This is to allow future developers to attempt} different merging and filtering strategies when compiling the input {\tt AstroRV} catalog.  

\item The version of the {\tt AstroRV} catalog used in this work is also provided with the package.  As a result, the full size of the local copy is almost 3 GB in size.  A smaller size version of the package with a size of only 450 MB is available at \url{http://github.com/seap-udea/iWander}.

\item Any contribution to the development of the package is welcomed.  We can provide full access to the developing branch to any researcher or developer interested in to contribute with this project.  

\item The package, as well as the related databases will be updated as future and best astrometric and radial velocity information be published.  
\end{itemize}

\bibliography{bibliography} \bibliographystyle{apj2}

\begin{thebibliography}{70}
\expandafter\ifx\csname natexlab\endcsname\relax\def\natexlab#1{#1}\fi

\bibitem[{Acton~Jr(1996)}]{Acton1996}
Acton~Jr, C.~H. 1996, Planetary and Space Science, 44, 65

\bibitem[{{Adams} \& {Spergel}(2005)}]{Adams2005}
{Adams}, F.~C., \& {Spergel}, D.~N. 2005, Astrobiology, 5, 497

\bibitem[{Bailer-Jones(2017)}]{Bailer2017}
Bailer-Jones, C. 2017, Astronomy \& Astrophysics

\bibitem[{{Bailer-Jones}(2015)}]{Bailer2015}
{Bailer-Jones}, C.~A.~L. 2015, \aap, 575, A35

\bibitem[{Barbier-Brossat \& Figon(2000{\natexlab{a}})}]{Barbier2000a}
Barbier-Brossat, M., \& Figon, P. 2000{\natexlab{a}}, Astronomy and
  Astrophysics Supplement Series, 142, 217

\bibitem[{Barbier-Brossat \& Figon(2000{\natexlab{b}})}]{Barbier2000b}
---. 2000{\natexlab{b}}, Astronomy and Astrophysics Supplement Series, 142, 217

\bibitem[{{Belbruno} {et~al.}(2012){Belbruno}, {Moro-Mart{\'{\i}}n},
  {Malhotra}, \& {Savransky}}]{Belbruno2012}
{Belbruno}, E., {Moro-Mart{\'{\i}}n}, A., {Malhotra}, R., \& {Savransky}, D.
  2012, Astrobiology, 12, 754

\bibitem[{{Berski} \& {Dybczy{\'n}ski}(2016)}]{Berski2016}
{Berski}, F., \& {Dybczy{\'n}ski}, P.~A. 2016, \aap, 595, L10

\bibitem[{{Bolin} {et~al.}(2018){Bolin}, {Weaver}, {Fernandez}, {Lisse},
  {Huppenkothen}, {Jones}, {Juri{\'c}}, {Moeyens}, {Schambeau}, {Slater},
  {Ivezi{\'c}}, \& {Connolly}}]{Bolin2017}
{Bolin}, B.~T., {Weaver}, H.~A., {Fernandez}, Y.~R., {et~al.} 2018, \apjl, 852,
  L2

\bibitem[{Bovy(2015)}]{Bovy2015}
Bovy, J. 2015, The Astrophysical Journal Supplement Series, 216, 29

\bibitem[{Brown {et~al.}(2016)Brown, Vallenari, Prusti, De~Bruijne, Mignard,
  Drimmel, Babusiaux, Bailer-Jones, Bastian, Biermann, {et~al.}}]{Brown2016}
Brown, A.~G., Vallenari, A., Prusti, T., {et~al.} 2016, Astronomy \&
  Astrophysics, 595, A2

\bibitem[{Bulirsch \& Stoer(1966)}]{Bulirsch1966}
Bulirsch, R., \& Stoer, J. 1966, Numerische Mathematik, 8, 1

\bibitem[{Casagrande {et~al.}(2011)Casagrande, Schoenrich, Asplund, Cassisi,
  Ramirez, Melendez, Bensby, \& Feltzing}]{Casagrande2011}
Casagrande, L., Schoenrich, R., Asplund, M., {et~al.} 2011, Astronomy \&
  Astrophysics, 530, A138

\bibitem[{{Chambers} {et~al.}(2016){Chambers}, {Magnier}, {Metcalfe},
  {Flewelling}, {Huber}, {Waters}, {Denneau}, {Draper}, {Farrow}, {Finkbeiner},
  {Holmberg}, {Koppenhoefer}, {Price}, {Saglia}, {Schlafly}, {Smartt},
  {Sweeney}, {Wainscoat}, {Burgett}, {Grav}, {Heasley}, {Hodapp}, {Jedicke},
  {Kaiser}, {Kudritzki}, {Luppino}, {Lupton}, {Monet}, {Morgan}, {Onaka},
  {Stubbs}, {Tonry}, {Banados}, {Bell}, {Bender}, {Bernard}, {Botticella},
  {Casertano}, {Chastel}, {Chen}, {Chen}, {Cole}, {Deacon}, {Frenk},
  {Fitzsimmons}, {Gezari}, {Goessl}, {Goggia}, {Goldman}, {Grebel}, {Hambly},
  {Hasinger}, {Heavens}, {Heckman}, {Henderson}, {Henning}, {Holman}, {Hopp},
  {Ip}, {Isani}, {Keyes}, {Koekemoer}, {Kotak}, {Long}, {Lucey}, {Liu},
  {Martin}, {McLean}, {Morganson}, {Murphy}, {Nieto-Santisteban}, {Norberg},
  {Peacock}, {Pier}, {Postman}, {Primak}, {Rae}, {Rest}, {Riess}, {Riffeser},
  {Rix}, {Roser}, {Schilbach}, {Schultz}, {Scolnic}, {Szalay}, {Seitz},
  {Shiao}, {Small}, {Smith}, {Soderblom}, {Taylor}, {Thakar}, {Thiel},
  {Thilker}, {Urata}, {Valenti}, {Walter}, {Watters}, {Werner}, {White},
  {Wood-Vasey}, \& {Wyse}}]{Chambers2016}
{Chambers}, K.~C., {Magnier}, E.~A., {Metcalfe}, N., {et~al.} 2016, ArXiv
  e-prints

\bibitem[{{{\'C}uk}(2018)}]{Matija2018}
{{\'C}uk}, M. 2018, \apjl, 852, L15

\bibitem[{Cvetkovi{\'c}(2011)}]{Cvetkovic2011}
Cvetkovi{\'c}, Z. 2011, The Astronomical Journal, 141, 116

\bibitem[{{de la Fuente Marcos} \& {de la Fuente
  Marcos}(2017)}]{delafuente2017}
{de la Fuente Marcos}, C., \& {de la Fuente Marcos}, R. 2017, Research Notes of
  the American Astronomical Society, 1, 5

\bibitem[{{Dybczy{\'n}ski} \& {Kr{\'o}likowska}(2017)}]{Dybczyski2017}
{Dybczy{\'n}ski}, P.~A., \& {Kr{\'o}likowska}, M. 2017, ArXiv e-prints

\bibitem[{{Engelhardt} {et~al.}(2017){Engelhardt}, {Jedicke}, {Vere{\v s}},
  {Fitzsimmons}, {Denneau}, {Beshore}, \& {Meinke}}]{Engelhardt2017}
{Engelhardt}, T., {Jedicke}, R., {Vere{\v s}}, P., {et~al.} 2017, \aj, 153, 133

\bibitem[{{ESA}(1997)}]{ESA1997}
{ESA}, ed. 1997, ESA Special Publication, Vol. 1200, {The HIPPARCOS and TYCHO
  catalogues. Astrometric and photometric star catalogues derived from the ESA
  HIPPARCOS Space Astrometry Mission}

\bibitem[{{Famaey} {et~al.}(2005){Famaey}, {Jorissen}, {Luri}, {Mayor}, {Udry},
  {Dejonghe}, \& {Turon}}]{Famaey2005}
{Famaey}, B., {Jorissen}, A., {Luri}, X., {et~al.} 2005, \aap, 430, 165

\bibitem[{{Feng} \& {Jones}(2018)}]{Feng2018}
{Feng}, F., \& {Jones}, H.~R.~A. 2018, \apjl, 852, L27

\bibitem[{{Ferrin} \& {Zuluaga}(2017)}]{Ferrin2017}
{Ferrin}, I., \& {Zuluaga}, J. 2017, ArXiv e-prints

\bibitem[{{Gaidos} {et~al.}(2017){Gaidos}, {Williams}, \& {Kraus}}]{Gaidos2017}
{Gaidos}, E., {Williams}, J., \& {Kraus}, A. 2017, Research Notes of the
  American Astronomical Society, 1, 13

\bibitem[{Galassi {et~al.}(2002)Galassi, Davies, Theiler, Gough, Jungman,
  Alken, Booth, \& Rossi}]{Galassi2002}
Galassi, M., Davies, J., Theiler, J., {et~al.} 2002, Network Theory Ltd, 3

\bibitem[{{Garc{\'{\i}}a-S{\'a}nchez}
  {et~al.}(2001){Garc{\'{\i}}a-S{\'a}nchez}, {Weissman}, {Preston}, {Jones},
  {Lestrade}, {Latham}, {Stefanik}, \& {Paredes}}]{Garcia2001}
{Garc{\'{\i}}a-S{\'a}nchez}, J., {Weissman}, P.~R., {Preston}, R.~A., {et~al.}
  2001, \aap, 379, 634

\bibitem[{{Gontcharov}(2006)}]{Gontcharov2006}
{Gontcharov}, G.~A. 2006, Astronomy Letters, 32, 759

\bibitem[{Gragg(1965)}]{Gragg1965}
Gragg, W.~B. 1965, Journal of the Society for Industrial and Applied
  Mathematics, Series B: Numerical Analysis, 2, 384

\bibitem[{Higuchi \& Ida(2017)}]{Higuchi2017}
Higuchi, A., \& Ida, S. 2017, The Astronomical Journal, 154, 88

\bibitem[{{H{\o}g} {et~al.}(2000){H{\o}g}, {Fabricius}, {Makarov}, {Urban},
  {Corbin}, {Wycoff}, {Bastian}, {Schwekendiek}, \& {Wicenec}}]{Hog2000}
{H{\o}g}, E., {Fabricius}, C., {Makarov}, V.~V., {et~al.} 2000, \aap, 355, L27

\bibitem[{Holmberg {et~al.}(2007)Holmberg, Nordstr{\"o}m, \&
  Andersen}]{Holmberg2007}
Holmberg, J., Nordstr{\"o}m, B., \& Andersen, J. 2007, Astronomy \&
  Astrophysics, 475, 519

\bibitem[{Hunter {et~al.}(2007)}]{Hunter2007}
Hunter, J.~D., {et~al.} 2007, Computing in science and engineering, 9, 90

\bibitem[{{Ivezic} {et~al.}(2008){Ivezic}, {Tyson}, {Abel}, {Acosta},
  {Allsman}, {AlSayyad}, {Anderson}, {Andrew}, {Angel}, {Angeli}, {Ansari},
  {Antilogus}, {Arndt}, {Astier}, {Aubourg}, {Axelrod}, {Bard}, {Barr},
  {Barrau}, {Bartlett}, {Bauman}, {Beaumont}, {Becker}, {Becla}, {Beldica},
  {Bellavia}, {Blanc}, {Blandford}, {Bloom}, {Bogart}, {Borne}, {Bosch},
  {Boutigny}, {Brandt}, {Brown}, {Bullock}, {Burchat}, {Burke}, {Cagnoli},
  {Calabrese}, {Chandrasekharan}, {Chesley}, {Cheu}, {Chiang}, {Claver},
  {Connolly}, {Cook}, {Cooray}, {Covey}, {Cribbs}, {Cui}, {Cutri}, {Daubard},
  {Daues}, {Delgado}, {Digel}, {Doherty}, {Dubois}, {Dubois-Felsmann},
  {Durech}, {Eracleous}, {Ferguson}, {Frank}, {Freemon}, {Gangler}, {Gawiser},
  {Geary}, {Gee}, {Geha}, {Gibson}, {Gilmore}, {Glanzman}, {Goodenow},
  {Gressler}, {Gris}, {Guyonnet}, {Hascall}, {Haupt}, {Hernandez}, {Hogan},
  {Huang}, {Huffer}, {Innes}, {Jacoby}, {Jain}, {Jee}, {Jernigan},
  {Jevremovic}, {Johns}, {Jones}, {Juramy-Gilles}, {Juric}, {Kahn}, {Kalirai},
  {Kallivayalil}, {Kalmbach}, {Kantor}, {Kasliwal}, {Kessler}, {Kirkby},
  {Knox}, {Kotov}, {Krabbendam}, {Krughoff}, {Kubanek}, {Kuczewski},
  {Kulkarni}, {Lambert}, {Le Guillou}, {Levine}, {Liang}, {Lim}, {Lintott},
  {Lupton}, {Mahabal}, {Marshall}, {Marshall}, {May}, {McKercher}, {Migliore},
  {Miller}, {Mills}, {Monet}, {Moniez}, {Neill}, {Nief}, {Nomerotski},
  {Nordby}, {O'Connor}, {Oliver}, {Olivier}, {Olsen}, {Ortiz}, {Owen}, {Pain},
  {Peterson}, {Petry}, {Pierfederici}, {Pietrowicz}, {Pike}, {Pinto}, {Plante},
  {Plate}, {Price}, {Prouza}, {Radeka}, {Rajagopal}, {Rasmussen}, {Regnault},
  {Ridgway}, {Ritz}, {Rosing}, {Roucelle}, {Rumore}, {Russo}, {Saha},
  {Sassolas}, {Schalk}, {Schindler}, {Schneider}, {Schumacher}, {Sebag},
  {Sembroski}, {Seppala}, {Shipsey}, {Silvestri}, {Smith}, {Smith}, {Strauss},
  {Stubbs}, {Sweeney}, {Szalay}, {Takacs}, {Thaler}, {Van Berg}, {Vanden Berk},
  {Vetter}, {Virieux}, {Xin}, {Walkowicz}, {Walter}, {Wang}, {Warner},
  {Willman}, {Wittman}, {Wolff}, {Wood-Vasey}, {Yoachim}, {Zhan}, \& {for the
  LSST Collaboration}}]{Ivezic2008}
{Ivezic}, Z., {Tyson}, J.~A., {Abel}, B., {et~al.} 2008, ArXiv e-prints

\bibitem[{{Jewitt} {et~al.}(2017){Jewitt}, {Luu}, {Rajagopal}, {Kotulla},
  {Ridgway}, {Liu}, \& {Augusteijn}}]{Jewitt2017}
{Jewitt}, D., {Luu}, J., {Rajagopal}, J., {et~al.} 2017, \apjl, 850, L36

\bibitem[{Johnson \& Soderblom(1987)}]{Johnson1987}
Johnson, D.~R., \& Soderblom, D.~R. 1987, The Astronomical Journal, 93, 864

\bibitem[{Karim \& Mamajek(2016)}]{Karim2016}
Karim, M.~T., \& Mamajek, E.~E. 2016, Monthly Notices of the Royal Astronomical
  Society, stw2772

\bibitem[{Kolmogorov(1968)}]{Kolmogorov1968}
Kolmogorov, A.~N. 1968, Foundations of the Theory of Probability (Courier Dover
  Publications)

\bibitem[{Kunder {et~al.}(2017)Kunder, Kordopatis, Steinmetz, Zwitter,
  McMillan, Casagrande, Enke, Wojno, Valentini, Chiappini,
  {et~al.}}]{Kunder2017}
Kunder, A., Kordopatis, G., Steinmetz, M., {et~al.} 2017, The Astronomical
  Journal, 153, 75

\bibitem[{Kuzmin(1956)}]{Kuzmin1956}
Kuzmin, G. 1956, Astronomicheskii Zhurnal, 33, 27

\bibitem[{{Malaroda} {et~al.}(2000){Malaroda}, {Levato}, {Morrell},
  {Garc{\'{\i}}a}, {Grosso}, \& {Bolzicco}}]{Malaroda2000}
{Malaroda}, S., {Levato}, H., {Morrell}, N., {et~al.} 2000, \aaps, 144, 1

\bibitem[{{Maldonado} {et~al.}(2010){Maldonado}, {Mart{\'{\i}}nez-Arn{\'a}iz},
  {Eiroa}, {Montes}, \& {Montesinos}}]{Maldonado2010}
{Maldonado}, J., {Mart{\'{\i}}nez-Arn{\'a}iz}, R.~M., {Eiroa}, C., {Montes},
  D., \& {Montesinos}, B. 2010, \aap, 521, A12

\bibitem[{{Mamajek}(2017)}]{Mamajek2017}
{Mamajek}, E. 2017, Research Notes of the American Astronomical Society, 1, 21

\bibitem[{Martell {et~al.}(2016)Martell, Sharma, Buder, Duong, Schlesinger,
  Simpson, Lind, Ness, Marshall, Asplund, {et~al.}}]{Martell2016}
Martell, S., Sharma, S., Buder, S., {et~al.} 2016, Monthly Notices of the Royal
  Astronomical Society, stw2835

\bibitem[{{Masiero}(2017)}]{Masiero2017}
{Masiero}, J. 2017, ArXiv e-prints

\bibitem[{{McGlynn} \& {Chapman}(1989)}]{McGlynn1989}
{McGlynn}, T.~A., \& {Chapman}, R.~D. 1989, \apjl, 346, L105

\bibitem[{{Melosh}(2003)}]{Melosh2003}
{Melosh}, H.~J. 2003, Astrobiology, 3, 207

\bibitem[{{Merlin} {et~al.}(2017){Merlin}, {Hromakina}, {Perna}, {Hong}, \&
  {Alvarez-Candal}}]{Merlin2017}
{Merlin}, F., {Hromakina}, T., {Perna}, D., {Hong}, M.~J., \& {Alvarez-Candal},
  A. 2017, \aap, 604, A86

\bibitem[{Miyamoto \& Nagai(1975)}]{Miyamoto1975}
Miyamoto, M., \& Nagai, R. 1975, Publications of the Astronomical Society of
  Japan, 27, 533

\bibitem[{{Napier}(2004)}]{Napier2004}
{Napier}, W.~M. 2004, \mnras, 348, 46

\bibitem[{{Opik}(1932)}]{Opik1932}
{Opik}, E. 1932, The Scientific Monthly, 35, 109

\bibitem[{P{\'e}rez \& Granger(2007)}]{Perez2007}
P{\'e}rez, F., \& Granger, B.~E. 2007, Computing in Science \& Engineering, 9,
  21

\bibitem[{{Perryman} {et~al.}(1997){Perryman}, {Lindegren}, {Kovalevsky},
  {Hoeg}, {Bastian}, {Bernacca}, {Cr{\'e}z{\'e}}, {Donati}, {Grenon},
  {Grewing}, {van Leeuwen}, {van der Marel}, {Mignard}, {Murray}, {Le Poole},
  {Schrijver}, {Turon}, {Arenou}, {Froeschl{\'e}}, \&
  {Petersen}}]{Perryman1997}
{Perryman}, M.~A.~C., {Lindegren}, L., {Kovalevsky}, J., {et~al.} 1997, \aap,
  323, L49

\bibitem[{{Portegies Zwart} {et~al.}(2017){Portegies Zwart}, {Pelupessy},
  {Bedorf}, {Cai}, \& {Torres}}]{Portegies2017}
{Portegies Zwart}, S., {Pelupessy}, I., {Bedorf}, J., {Cai}, M., \& {Torres},
  S. 2017, ArXiv e-prints

\bibitem[{Price(2012)}]{Price2012}
Price, D.~J. 2012, Journal of Computational Physics, 231, 759

\bibitem[{{Raymond} {et~al.}(2017){Raymond}, {Armitage}, {Veras}, {Quintana},
  \& {Barclay}}]{Raymond2017}
{Raymond}, S.~N., {Armitage}, P.~J., {Veras}, D., {Quintana}, E.~V., \&
  {Barclay}, T. 2017, ArXiv e-prints

\bibitem[{Sch{\"o}nrich {et~al.}(2010)Sch{\"o}nrich, Binney, \&
  Dehnen}]{Schonrich2010}
Sch{\"o}nrich, R., Binney, J., \& Dehnen, W. 2010, Monthly Notices of the Royal
  Astronomical Society, 403, 1829

\bibitem[{{Sen} \& {Rama}(1993)}]{Sen1993}
{Sen}, A.~K., \& {Rama}, N.~C. 1993, \aap, 275, 298

\bibitem[{{Strom} {et~al.}(2005){Strom}, {Malhotra}, {Ito}, {Yoshida}, \&
  {Kring}}]{Strom2005}
{Strom}, R.~G., {Malhotra}, R., {Ito}, T., {Yoshida}, F., \& {Kring}, D.~A.
  2005, Science, 309, 1847

\bibitem[{{Strom Robert} {et~al.}(2015){Strom Robert}, {Renu}, {Zhi-Yong},
  {Takashi}, {Fumi}, \& {Ostrach Lillian}}]{Strom2015}
{Strom Robert}, G., {Renu}, M., {Zhi-Yong}, X., {et~al.} 2015, Research in
  Astronomy and Astrophysics, 15, 407

\bibitem[{Thebault(2011)}]{Thebault2011}
Thebault, P. 2011, Celestial Mechanics and Dynamical Astronomy, 111, 29

\bibitem[{{Trilling} {et~al.}(2017){Trilling}, {Robinson}, {Roegge},
  {Chandler}, {Smith}, {Loeffler}, {Trujillo}, {Navarro-Meza}, \&
  {Glaspie}}]{Trilling2017}
{Trilling}, D.~E., {Robinson}, T., {Roegge}, A., {et~al.} 2017, \apjl, 850, L38

\bibitem[{Van Der~Walt {et~al.}(2011)Van Der~Walt, Colbert, \&
  Varoquaux}]{Van2011}
Van Der~Walt, S., Colbert, S.~C., \& Varoquaux, G. 2011, Computing in Science
  \& Engineering, 13, 22

\bibitem[{{Wenger} {et~al.}(2000){Wenger}, {Ochsenbein}, {Egret}, {Dubois},
  {Bonnarel}, {Borde}, {Genova}, {Jasniewicz}, {Lalo{\"e}}, {Lesteven}, \&
  {Monier}}]{Wenger2000}
{Wenger}, M., {Ochsenbein}, F., {Egret}, D., {et~al.} 2000, \aaps, 143, 9

\bibitem[{{Wiegert}(2011)}]{Wiegert2011}
{Wiegert}, P.~A. 2011, in Meteoroids: The Smallest Solar System Bodies, ed.
  W.~J. {Cooke}, D.~E. {Moser}, B.~F. {Hardin}, \& D.~{Janches}, 106

\bibitem[{{Wiegert}(2014)}]{Wiegert2014}
{Wiegert}, P.~A. 2014, \icarus, 242, 112

\bibitem[{{Ye} {et~al.}(2017){Ye}, {Zhang}, {Kelley}, \& {Brown}}]{Ye2017}
{Ye}, Q.-Z., {Zhang}, Q., {Kelley}, M.~S.~P., \& {Brown}, P.~G. 2017, \apjl,
  851, L5

\bibitem[{Zhang(2018)}]{Zhang2018}
Zhang, Q. 2018, The Astrophysical Journal Letters, 852, L13

\bibitem[{Zucker {et~al.}(2004)Zucker, Mazeh, Santos, Udry, \&
  Mayor}]{Zucker2004}
Zucker, S., Mazeh, T., Santos, N., Udry, S., \& Mayor, M. 2004, Astronomy \&
  Astrophysics, 426, 695

\bibitem[{{Zuluaga} {et~al.}(2013){Zuluaga}, {Ferrin}, \&
  {Geens}}]{Zuluaga2013}
{Zuluaga}, J.~I., {Ferrin}, I., \& {Geens}, S. 2013, ArXiv e-prints

\bibitem[{Zuluaga \& Sucerquia(2018)}]{Zuluaga2018}
Zuluaga, J.~I., \& Sucerquia, M. 2018, Monthly Notices of the Royal
  Astronomical Society, sty702

\end{thebibliography}


\end{document}